\newif\ifmain
\newif\ifappendix
\newif\iffull
\newif\ifdemo
\newif\ifdemoext

\fulltrue

\ifmain
\def\papertitle{Searching For Music Mixing Graphs: A Pruning Approach}
\fi
\ifappendix
\def\papertitle{Searching For Music Mixing Graphs: A Pruning Approach}
\fi
\iffull
\def\papertitle{Searching For Music Mixing Graphs: A Pruning Approach}
\fi

\def\paperauthorA{Sungho Lee}
\def\paperauthorB{Marco A. Martínez-Ramírez}
\def\paperauthorC{Wei-Hsiang Liao}
\def\paperauthorD{Stefan Uhlich}
\def\paperauthorE{Giorgio Fabbro}
\def\paperauthorF{Kyogu Lee}
\def\paperauthorG{Yuki Mitsufuji}

\documentclass[twoside,a4paper]{article}

\usepackage{dafx_24}
\usepackage{amsmath,amssymb,amsfonts,amsthm}
\usepackage{etoolbox}
\usepackage{euscript}
\usepackage[T1]{fontenc}
\usepackage[utf8]{inputenc}
\usepackage{ifpdf}
\usepackage[english]{babel}
\usepackage{stfloats}

\usepackage{subfig} 
\usepackage{caption}
\usepackage{xcolor}
\usepackage{booktabs}
\usepackage{colortbl}
\usepackage{multirow}
\usepackage{enumitem}
\usepackage{siunitx}
\usepackage{musicography}
\usepackage{fancyvrb}
\usepackage{listings}
\definecolor{codegreen}{rgb}{0,0.6,0}
\definecolor{codegray}{rgb}{0.5,0.5,0.5}
\definecolor{codepurple}{rgb}{0.58,0,0.82}
\definecolor{backcolour}{rgb}{0.96,0.96,0.92}

\makeatletter
\lst@AddToHook{OnEmptyLine}{\addtocounter{lstnumber}{-1}}
\makeatother

\lstdefinestyle{mystyle}{
    backgroundcolor=\color{backcolour},   
    commentstyle=\color{codegreen},
    keywordstyle=\color{magenta},
    numberstyle=\tiny\color{codegray},
    stringstyle=\color{codepurple},
    basicstyle=\ttfamily\footnotesize,
    breakatwhitespace=false,         
    breaklines=true,                 
    captionpos=b,                    
    keepspaces=true,                 
    numbers=left,   
    numberstyle=\ttfamily\scriptsize,
    numbersep=5pt,                  
    showspaces=false,                
    numberblanklines=false,
    showstringspaces=false,
    showtabs=false,                  
    tabsize=2,
        literate={0}{{\textcolor{blue}{0}}}{1}%
             {1}{{\textcolor{blue}{1}}}{1}%
             {2}{{\textcolor{blue}{2}}}{1}%
             {3}{{\textcolor{blue}{3}}}{1}%
             {4}{{\textcolor{blue}{4}}}{1}%
             {5}{{\textcolor{blue}{5}}}{1}%
             {6}{{\textcolor{blue}{6}}}{1}%
             {7}{{\textcolor{blue}{7}}}{1}%
             {8}{{\textcolor{blue}{8}}}{1}%
             {9}{{\textcolor{blue}{9}}}{1}%
}

\DeclareMathOperator*{\argmin}{arg\,min}

\usepackage{algorithm}
\usepackage{algpseudocode}
\algrenewcommand\algorithmicrequire{\textbf{Input:}}
\algrenewcommand\algorithmicensure{\textbf{Output:}}
\algnewcommand\algorithmicinput{\textbf{Input:}}
\algnewcommand\algorithmicoutput{\textbf{Output:}}
\algnewcommand\Input{\item[\algorithmicinput]}%
\algnewcommand\Output{\item[\algorithmicoutput]}%

\renewcommand\paragraph[1]{\noindent\textbf{#1 ---}}
\newcommand\Tc{T_\mathrm{pool}}

\newcommand\bfS{\mathbf{S}}

\newcommand\bfP{\mathbf{P}}

\newcommand\Gc{G_\mathrm{c}}
\newcommand\Gp{G_\mathrm{p}}
\newcommand\Vc{V_\mathrm{cand}}

\newcommand\Vp{V_\mathrm{p}}
\newcommand\bfw{\mathbf{w}}
\newcommand\bfm{\mathbf{m}}

\newcommand\La{L_\mathrm{a}}
\newcommand\Lr{L_\mathrm{r}}
\newcommand\Lg{L_\mathrm{g}}
\newcommand\Lp{L_\mathrm{p}}
\newcommand\fn[1]{\mathrm{#1}\:\!}
\newcommand\textttt[1]{\text{\scriptsize$\texttt{#1}$}}

\newcolumntype{C}[1]{>{\centering\arraybackslash}p{#1}}

\input glyphtounicode
\ninept

\newcounter{numauth}
\newcounter{listcnt}
\newcommand\authcnt[1]{\ifdefined#1 \stepcounter{numauth} \fi}

\newcommand\addauth[1]{
\ifdefined#1 
\stepcounter{listcnt}
\ifnum \value{listcnt}<\value{numauth}
\appto\authorslist{, #1}
\else
\appto\authorslist{~and~#1}
\fi
\fi}
\authcnt{\paperauthorB}
\authcnt{\paperauthorC}
\authcnt{\paperauthorD}
\authcnt{\paperauthorE}
\authcnt{\paperauthorF}
\authcnt{\paperauthorG}
\authcnt{\paperauthorH}
\authcnt{\paperauthorI}
\authcnt{\paperauthorJ}
\def\authorslist{\paperauthorA}
\addauth{\paperauthorB}
\addauth{\paperauthorC}
\addauth{\paperauthorD}
\addauth{\paperauthorE}
\addauth{\paperauthorF}
\addauth{\paperauthorG}
\addauth{\paperauthorH}
\addauth{\paperauthorI}
\addauth{\paperauthorJ}

\usepackage{times}

\newif\ifpdf
\ifx\pdfoutput\relax
\else
\ifcase\pdfoutput
\pdffalse
\else
\pdftrue
\fi

\ifpdf 
\usepackage[pdftex,
pdftitle={\papertitle},
pdfauthor={\authorslist},
pdfsubject={Proceedings of the 27th International Conference on Digital Audio Effects (DAFx24)},
colorlinks=false, 
bookmarksnumbered, 
pdfstartview=XYZ 
]{hyperref}
\usepackage[pdftex]{graphicx}
\else 
\usepackage[dvips]{epsfig,graphicx}
\usepackage[dvips,
pdftitle={\papertitle},
pdfauthor={\authorslist},
pdfsubject={Proceedings of the 27th International Conference on Digital Audio Effects (DAFx24)},
colorlinks=false, 
bookmarksnumbered, 
pdfstartview=XYZ 
]{hyperref}
\fi
\usepackage[hypcap=true]{caption}
\usepackage[edges]{forest}

\title{\papertitle}

\affiliation{
\hspace{-4.5mm}
\ifdemo
\paperauthorA$^\dagger$\sthanks{Work partially done during an internship at Sony AI. 
},\, 
\fi
\ifdemoext
\paperauthorA$^\dagger$\sthanks{Work partially done during an internship at Sony AI. 
},\, 
\fi
\iffull
\paperauthorA$^\dagger$\sthanks{Work done during an internship at Sony AI. 
},\, 
\fi
\ifmain
\paperauthorA$^\dagger$\sthanks{Work done during an internship at Sony AI. 
},\, 
\fi
\paperauthorB$^\natural$,\, \paperauthorC$^\natural$,\, \paperauthorD$^\sharp$,\, \paperauthorE$^\sharp$,\, \paperauthorF$^\dagger$,\, and \paperauthorG$^{\natural\flat}$
}
{
\hspace{-5mm}
$^\dagger$Department of Intelligence and Information, Seoul National University, Seoul, South Korea \\
\hspace{-5mm}
$^\natural$Sony AI, Tokyo, Japan \quad
$^\sharp$Sony Europe B.V., Stuttgart, Germany \quad 
$^\flat$Sony Group Corporation, Tokyo, Japan
}
\definecolor{emerald}{rgb}{0.31, 0.78, 0.47}

\usepackage{tikz}
\usetikzlibrary{external}
\usetikzlibrary{arrows.meta, bending, decorations.pathmorphing}

\begin{document}
\ifpdf 
\DeclareGraphicsExtensions{.png,.jpg,.pdf}
\else  
\DeclareGraphicsExtensions{.eps}
\fi

\makeatletter
\pdfbookmark[0]{\@pdftitle}{title}
\makeatother


\ifmain
\maketitle
\begin{abstract}
Music mixing is \emph{compositional} --- experts combine multiple audio processors to achieve a cohesive mix from dry source tracks. We propose a method to reverse engineer this process from the input and output audio. 
First, we create a mixing console that applies all available processors to every chain. Then, after the initial console parameter optimization, we alternate between removing redundant processors and fine-tuning. 
We achieve this through differentiable implementation of both processors and pruning.
Consequently, we find a sparse mixing graph that achieves nearly identical matching quality of the full mixing console. 
We apply this procedure to dry-mix pairs from various datasets and collect graphs that also can be used to train neural networks for music mixing applications.
\end{abstract}

\section{Introduction}
\paragraph{Motivation}
From a signal processing perspective, modern music is more than the mere sum of source tracks.
Mixing engineers combine and control multiple processors
to balance the sources in terms of loudness, frequency content, spatialization, and much more.
Many attempts have been made to uncover parts of this intricate process.
Some have gathered expert knowledge \cite{pestana2014intelligent, everardo2017towards}
and built rule-based systems \cite{perez2009automatic, de2013knowledge}. 
More recent work has adopted data-driven approaches. 
Neural networks have been trained to map source tracks directly to a mix \cite{martinez2022automatic, koszewski2023automatic}
or to estimate parameters of a fixed processing chain \cite{steinmetz2020diffmixconsole}.
Yet, efforts to address the compositional aspects of the music mixing,
such as which processors to use for each track, are still limited.
One possible remedy is to consider a graph representation whose nodes and edges are processors and connections between them, respectively.
In other words, each graph contains the essential information 
about the mixing process.
However, other than the dry source and mixed audio, no public dataset provides such mixing graphs or related metadata
\cite{bittner2014medleydb, bittner2016medleydb, senior2018mixing},
which hinders this line of research.
This is not surprising; besides the cost of crowdsourcing,
it is difficult to standardize the mixing data from multiple engineers
with different equipment.
One recent work \cite{lee2023blind} sidestepped this issue by creating synthetic graphs and using them for training.
However, this approach is not free from downsides.
Neural networks would suffer from poor generalization
unless the synthetic data distribution matches the real world.
Similar data-related issues arise in different domains, e.g., audio effect chain recognition \cite{Mitcheltree_2021, guo2023automatic}
and synthesizer sound matching \cite{masuda2023improving, uzrad2024diffmoog, caspe2022ddx7}.
Furthermore, real-world multitrack mixes have a much larger number of source tracks and graph sizes, making synthetic data generation more challenging. 
Therefore, it is desirable to have a systematic, reliable, and scalable method for collecting graphs.
All these contexts lead us to ask: 
\emph{Can we find the mixing graphs solely from audio?}

\begin{figure}
    \centering
    \begin{tikzpicture}[>=Triangle, dot/.style={circle,fill=black,inner sep=1pt}, xscale=1.1, yscale=1.1]
        \tikzset{decoration={snake,amplitude=.4mm,segment length=2mm, post length=1.5mm,pre length=0mm}}
        \fill[gray, thick, fill=gray!10] 
            (-4+0.2, -2+0.2) rectangle (0-0.2, 2-0.2);
        \node at (-4+0.5+.15, 2-0.5) (gset) {$\mathcal{G}, \mathcal{P}$};

        \draw[red,thick,fill=red!20,xshift=-2cm] 
            plot[smooth cycle] 
            coordinates {
                (10:1.5) (80:1.5) (140:1.6) (200:1.5) (250:1.5) (290:1.5) (310:1.5)
            };

        \draw[orange,thick,fill=orange!30,xshift=-2.2cm,yshift=.35cm] plot[smooth cycle]
        coordinates {(10:1.2) (90:.9) (160:1) (200:.9) (250:.5)};

        \fill[gray, thick, fill=gray!10] (0+0.2, -2+0.2) rectangle (4-0.2, 2-0.2);

        \draw[teal,thick,fill=teal!30] (3-0.2, 2-0.2) -- (0+0.2, 2-0.2) -- (0+0.2,-2+0.2) -- (3.4-0.3,-2+0.2) 
                        to [out=140,in=-110]  (3.7-0.3, 2-0.2) --cycle ;

        \begin{scope}
            \draw[clip, ] 
                (3-0.2, 2-0.2) -- (0+0.2, 2-0.2) -- (0+0.2,-2+0.2) -- (3.4-0.3,-2+0.2) 
                        to [out=140,in=-110]  (3.7-0.3, 2-0.2) --cycle;
            \fill [emerald!20] (2.8, -.1) circle (1.4);         
        \end{scope}

        \draw[teal,thick] (3-0.2, 2-0.2) -- (0+0.2, 2-0.2) -- (0+0.2,-2+0.2) -- (3.4-0.3,-2+0.2) 
                        to [out=140,in=-110]  (3.7-0.3, 2-0.2) --cycle ;

        \draw
        (-2.7,  0.2) node (gc){$\Gc$}
        (-2.25,  0.75) node[dot,label={above:$\mathbf{0}$}] (g0){}
        (-1.6,  0.4) node[dot] (g){}
        (-1.35, -1.15) node[dot,label={left:$G_\mathrm{p}, \bfP_\mathrm{p}$}] (gp){}

        (-1.48, 0.61) node () {$\bfP_\mathrm{c}$}
        (-1.1, -0.1) node[inner sep=0, scale=0.02] (gp1){}
        (-1.6, -0.3) node[inner sep=0, scale=0.02] (gp2){}
        (-1.2, -0.7) node[inner sep=0, scale=0.02] (gp3){}
        ( 1.2,  1.1) node[dot, label={above:$\hat{y}_0$}] (y0){}
        ( 2.45,  0.4) node[dot, label={below:$\hat{y}_\mathrm{c}$}] (yhat){}
        ( 1.98,  1.19) node (tau){}
        ( 2.0, -.9) node[dot, label={right:$\hat{y}_\mathrm{p}$}] (yhatp){}
        ( 2.8,  -0.2) node[dot,label={right:${y}$}] (y){}
        
        (1.85, 0.1) node[inner sep=0, scale=0.02] (yp1){}
        (2.1, -0.3) node[inner sep=0, scale=0.02] (yp2){}
        (1.6, -0.5) node[inner sep=0, scale=0.02] (yp3){}
        
        (2.38, 0.8) node () {$\tau$}
       
        ;
        
        \draw[blue,thick,decorate,-{Stealth[bend]}] (g) -- (gp1);
        \draw[blue,thick,decorate,-{Stealth[bend]}] (gp2) -- (gp3);
        \draw[thick,-{Stealth[bend]}] (gp1) -- (gp2);
        \draw[thick,-{Stealth[bend]}] (gp3) -- (gp);
        \draw[thick,-{Stealth[bend]}] (g0) -- (g);

        \draw[thick,darkgray,dotted] (gc) -- (g0);
        \draw[thick,darkgray,dotted] (gc) -- (g);
        \draw[thick,darkgray,dotted] (g0) -- (y0);

        \draw[thick,darkgray,dotted] (g) -- (yhat);
        \draw[thick,darkgray,dotted] (gp) -- (yhatp);
        \draw[thick,gray] (y) -- (yhat);
        \draw[thick,gray] (yhat) -- (tau);

        \draw[thick,-{Stealth[bend]}] (y0) -- (yhat);
        \draw[blue,thick,decorate,-{Stealth[bend]}] (yhat) -- (yp1);
        \draw[blue,thick,decorate,-{Stealth[bend]}] (yp2) -- (yp3);
        \draw[thick,-{Stealth[bend]}] (yp1) -- (yp2);
        \draw[thick,-{Stealth[bend]}] (yp3) -- (yhatp);

        \begin{scope}
            \draw[clip, draw=none] 
                    (0+0.2, -2+0.2) rectangle (4-0.2, 2-0.2);   
            \draw[thick, gray] (2.8, -.1) circle (1.4); 
        \end{scope}

        \draw
        (-.9-2-0.08+0.11, -2.2) node (dummy1) {}
        (-.9-1.2-0.08+0.11, -2.2) node (dummy2) {}
        (-0.7+0.11, -2.22) node () {\footnotesize $\!\!\!\mathrm{Parameter\: optimization}$}
        (-.9+1.7+0+0.11, -2.2) node (dummy3) {}
        (-.9+1.7+0.8+0.11, -2.2) node (dummy4) {}
        (2.15+0.11, -2.22) node () {\footnotesize $\!\!\!\mathrm{Pruning}$} ;

        \draw[blue,thick,decorate,-{Stealth[bend]}] (dummy3) -- (dummy4);
        \draw[thick,-{Stealth[bend]}] (dummy1) -- (dummy2);
	
        \node at (+0.5, 2-0.5) (yset) {$\mathcal{Y}$};

        \draw[orange,thick,fill=orange!30] 
            (-3.5+0.02, -2.65-0.016+0.01) rectangle (-3.5+0.2, -2.45-0.016-0.01);
        \draw[text width=3cm]
            (-3.5 + 1.85-0.016, -2.574-0.016) node () {\footnotesize $\!\!\!\mathrm{Console\: parameters}$};

        \draw[red,thick,fill=red!20] 
            (-0.6+0.02, -2.65-0.016+0.01) rectangle (-0.6+0.2, -2.45-0.016-0.01);
        \draw[text width=3cm]
            (-0.6 + 1.85, -2.574-0.016) node () {\footnotesize $\!\!\!\mathrm{Pruning\: search \: with \: parameters}$};

        \draw[teal,thick,fill=teal!30] 
            (-3.5+0.02, -2.65-0.325-0.011+0.01) rectangle (-3.5+0.2, -2.45-0.325-0.011-0.01);
        \draw[text width=3cm]
            (-3.5 + 1.85, -2.574-0.325-0.011) node () {\footnotesize $\!\!\!\mathrm{Audio\: search\: space}$};
        
        \fill[emerald,thick,fill=emerald!20] 
            (-0.6+0.02, -2.65-0.325-0.011+0.01) rectangle (-0.6+0.2, -2.45-0.325-0.011-0.01);
        \draw[gray,thick] 
            (-0.6+0.02, -2.65-0.325-0.011+0.01) rectangle (-0.6+0.2, -2.45-0.325-0.011-0.01);
        \draw[text width=3cm]
            (-0.6 + 1.85, -2.574-0.325-0.011) node () {\footnotesize $\!\!\!\mathrm{Tolerance\: threshold\: range}$};

        \end{tikzpicture}
    \vspace{-5.5mm}
    \caption{
    \it
    Music mixing graph search via iterative pruning. 
    }
    \label{fig:framework}
    \vspace{-2.3mm}
\end{figure}
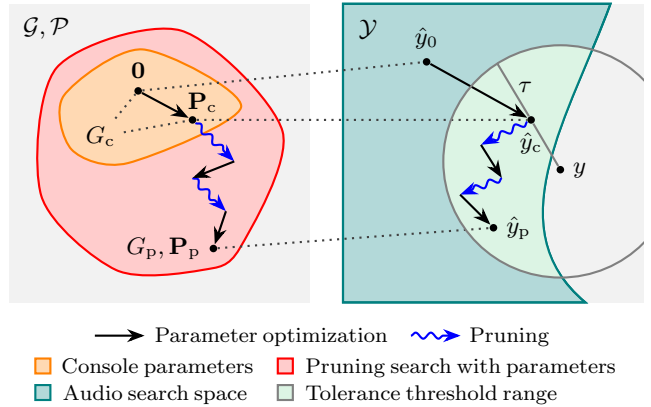

\vspace{2mm}
\paragraph{Problem definition}
Precisely, for each song (piece) whose dry sources $s_1, \cdots, s_K$ and mix $y$ are available, we aim to find an audio processing graph $G$ and its processor parameters $\bfP$ so that processing the dry sources $s_1, \cdots, s_K$ results in a mix $\hat{y}$ that closely matches the original mix $y$.
With a loss $\La$ that measures the match quality on the mixture audio domain $\mathcal{Y}$ and regularization $\Lr$,
our objective can be written as follows,
\begin{equation} \label{eq:objective}
        G^*, \bfP^* = 
        \argmin_{G, \bfP}
        \big[ L_\mathrm{a} (\hat{y}, y) + L_\mathrm{r} (G, \bfP) \big].
\end{equation} 

\paragraph{Contributions}
\begin{figure*}
    \centering
    \captionsetup[subfloat]{captionskip=5pt}
    \subfloat[Full mixing console (before pruning) \label{fig:full-graph}]{
        \includegraphics[height=2.47cm]{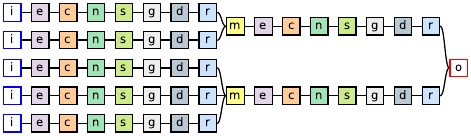}} \hspace{2mm}
    \subfloat[Pruned graph \label{fig:sparse-graph}]{
        \includegraphics[height=2.47cm]{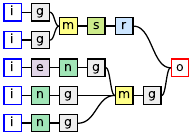}}\\
\vspace{-1mm}
\caption{
\it
	Finding the sparse graph $G_\mathrm{p}$ from the differentiable mixing console $G_\mathrm{c}$. 
    Initial letters in the nodes denote their respective types.
    \texttt{i}: input, \texttt{o}: output, \texttt{m}: mix, \texttt{e}: equalizer, \texttt{c}: compressor, \texttt{n}: noisegate, \texttt{s}: stereo imager, \texttt{g}: gain/panning,
\texttt{r}: reverb, \texttt{d}: multitap delay.
    }
  \label{fig:pruning} 
\vspace{-2mm}
\end{figure*}

One might want to explore the candidate graphs without any restriction. However, this makes the problem ill-posed and underdetermined. The graph's combinatorial nature makes the search space $\mathcal{G}$ extremely large. 
Furthermore, we have to find the processor parameters jointly. 
As a result, numerous pairs of graphs and parameters can have similar match quality.
Therefore, it is desirable to add some restrictions, e.g., preferring structures that are widely used by practitioners. 
To this end, we resort to the following pruning-based search; see Figure \ref{fig:framework} for a visual illustration.
Inspired by a recent work \cite{colonel2023music}, we first create a so-called ``mixing console'' $G_\mathrm{c}$ (see Figure \ref{fig:full-graph} for an example).
It applies a fixed processing chain to each source. Then, it subgroups the outputs, applies the chain again, and sums the processed subgroups to obtain a final mix $\hat{y}$.
This resembles the traditional hybrid mixing console \cite{hybridconsole}. 
Each chain comprises $7$ processors, including an equalizer, compressor, and multitap delay. 
We implement all of them in a differentiable manner \cite{engel2020ddsp, nercessian2020neural, steinmetz2022style}.
This allows end-to-end optimization of all parameters $\bfP_\mathrm{c}$ with an audio-domain loss $\La$ via gradient descent.
After this initial console training, we proceed to the pruning stage. 
Here, we search for a maximally pruned graph $G_\mathrm{p}$ and its parameters $\bfP_\mathrm{p}$ while maintaining the match quality of the mixing console up to a certain tolerance $\tau$; this is shown as a circle centered at $y$ in Figure \ref{fig:framework}. 
Also, see Figure \ref{fig:sparse-graph} for an example pruned graph.
We use iterative pruning, alternating between the pruning and fine-tuning, i.e., optimization of the remaining parameters \cite{castellano1997iterative}. 
To collect graphs from multiple songs, it is crucial to make the entire search efficient and fast. 
Pruning, in particular, takes a significant amount of computation time; hence, we investigate efficient and effective methods for pruning.
During the pruning, we need to find a subset of nodes that can be removed while not harming the match quality. 
To achieve this, we view each processor's “dry/wet” parameter as an approximate importance score and use it to select the candidate nodes. 
This approach gives $3$ variants of the pruning method with different trade-offs between the computational cost and resulting sparsity. 
It also draws connections to neural network pruning \cite{he2023structured, cheng2023survey} where the binary pruning operation is relaxed to continuous weights. 
Note that casting the graph search into pruning is a double-edged sword. 
The pruning only removes the processors and does not consider all possible signal routings, reducing the search space (from grey to colored regions in Figure \ref{fig:framework}). Consequently, it does not improve the match quality over the mixing consoles. 
Nevertheless, the pruned graph follows the real-world practice of selectively applying appropriate processors. 
In other words, the sparsity is crucial for the graph's interpretability. 
Also, it keeps the search cost in a practical range, which might be challenging with other alternatives \cite{liu2018darts, ye2023fm}.
Our method serves as a standalone reverse engineering algorithm \cite{colonel2023music}, but it can also be used to collect pseudo-label data to train neural networks for music mixing applications. For example, we may extend existing methods for automatic mixing \cite{perez2009automatic, de2013knowledge, 
martinez2022automatic, koszewski2023automatic, steinmetz2020diffmixconsole, steinmetz2022automix} and mixing style transfer \cite{koo2023music} to output the graphs. This allows end users to interpret and control the estimated outputs. 

\vspace{1.5mm}
\paragraph{Data}
We first report a list of datasets to which we can apply our method.
For each song, we need a pair of dry sources $s_1, \cdots, s_K$ and a final mixture $y$. 
Additionally, we use subgrouping information that describes how dry tracks are grouped together. 
Therefore, we use the \texttt{MedleyDB} dataset \cite{bittner2014medleydb, bittner2016medleydb} as it provides all of them. 
We also add the \texttt{MixingSecrets} library \cite{senior2018mixing}. 
Since it only provides the audio, we manually subgrouped each track based on its instrument.
Finally, we include our private dataset of Western music mixes from multiple engineers (denoted as \texttt{Internal}). 
The resulting ensemble comprises $1129$ songs ($188$, $472$, and $579$ songs for each respective dataset).
The number of dry tracks ranges from $1$ to $133$, and the number of subgroups ranges from $1$ to $26$ (see Figure \ref{fig:pruned-stats} for the statistics). 
Except for the final pruned graph collection stage (Section \ref{section:full-pruning}), 
we use a random subset for the evaluations (a total of $72$  songs, $24$ songs for each dataset). 
Every signal is stereo and resampled to $30\si{kHz}$ sampling rate. 

\paragraph{Supplementary materials} Refer to the following link for audio samples, pruned graphs, code, and appendices with supplementary details: 
\href{https://sh-lee97.github.io/grafx-prune}{\texttt{https://sh-lee97.github.io/grafx-prune}}.
\vspace{-3.5mm}
\section{Differentiable Processing on Graphs}
An audio processing graph $G=(V, E)$ is assumed to be directed and acyclic ($V$ and $E$ denote the set of nodes and edges, respectively).
Each node $v_i \in V$ is either a processor or an auxiliary module and has its type $t_i$, e.g., $\texttt{e}$ for an equalizer.
Each processor takes an audio $u_i$ and a parameter vector $p_i$ as input and outputs a processed signal $f_i(u_i, p_i)$.
Then, we further mix the input and this processed result with a ``dry/wet'' weight $w_i \in [0, 1]$.
Hence, the output $y_i$ of the processor $v_i$ is given as follows,
\begin{equation}\label{eq:computation}
    y_i =  w_i f_i(u_i, p_i) + (1-w_i)u_i.
\end{equation}
We have the following $3$ auxiliary modules: 
\begin{itemize}[leftmargin=4mm]
  \setlength\itemsep{0em}
    \item \textbf{Input ---} It outputs one of the dry sources $s_k$.
    \item \textbf{Mix ---} We output the sum of the incomming signals.
    \item \textbf{Output ---} A sum of its inputs is considered as a final output $y$.
\end{itemize}
Each edge $e_{ij} \in E$ represents a ``cable'' 
that sends an output signal to another node as input.
Throughout the text, we denote an ordered collection from multiple nodes with a boldface letter, 
e.g., $\bfw$ for a weight vector, $\bfS$ for a source tensor,
and $\bfP$ for a dictionary with processor types as keys and their parameter tensors as values.
Under this notation, our task is to find $G$, $\bfP$, and $\bfw$ from $\bfS$ and $y$.


\vspace{-.5mm}

\subsection{Differentiable Implementation}\label{section:diff_impl}
Considering the music mixing, 
we use the following $7$ processors. 

\begin{itemize}[leftmargin=4mm]
\setlength\itemsep{0em}
    \item \paragraph{Gain/panning} We control both loudness and stereo panning of input audio by multiplying a learnable scalar to each channel.

    \item \paragraph{Stereo imager} 
		We change the stereo width of the input by modifying the loudness of the side channel (left minus right).
    \item \paragraph{Equalizer} 
		We use a finite impulse response (FIR) with a length of $2047$ to modify the input's magnitude response.
            The FIR is parameterized with its log magnitude (thus $1024$ parameters).
            We apply inverse FFT of the magnitude with zero phase, obtain a zero-centered FIR, and multiply it with a Hann window.
            We apply the same FIR to both the left and right channels. 
        
    \item \paragraph{Reverb} 
		We employ $2$ seconds of filtered noise as an impulse response for reverberation. 
            First, we create a $2$-channel uniform noise, where these channels represent the mid and side.
            We filter the noise by multiplying an element-wise $2$-channel magnitude mask to its short-time Fourier transform (STFT), where the FFT sizes and hop lengths are $384$ and $192$, respectively.
            This mask is constructed using the reverberation's initial and decaying log magnitudes. 
            After the masking, we obtain the mid/side filtered noise via inverse STFT, convert it to stereo, and perform channel-wise convolutions with input to get an output.

    \item \paragraph{Compressor} 
            We use a slight variant of the recently proposed differentiable dynamic range compressor \cite{steinmetz2022style}.
            First, we obtain the input's smooth energy envelope. The smoothing is typically done with a ballistics filter, but we instead use a one-pole filter for speedup in GPU.
            Then, we compute the desired gain reduction from the envelope and apply it to the input audio.
            
    \item \paragraph{Noisegate} 
		Except for the gain computation, its implementation is the same as that of the compressor.
    \item \paragraph{Multitap delay} 
            For each (left and right) channel,
		we employ independent $2$ seconds of delay effects with a single delay for every $100\si{ms}$ interval.
            To optimize delay lengths using gradient descent, we employ surrogate complex damped sinusoids \cite{hayes2023sinusoidal}.
            Each sinusoid is converted to a delayed soft impulse via inverse FFT. 
            Its angular frequency represents a continuous relaxation of the discrete delay length.
            Each delay is filtered with a length-$39$ FIR equalizer to mimic the filtered echo effect \cite{zolzer2011dafx}.

\end{itemize}

\vspace{1.2mm}
\paragraph{Batched node processing}
It is common to compute the graph output signal by processing each node one by one \cite{uzrad2024diffmoog, engel2020ddsp}. However, this severely bottlenecks the computation speed for large mixing graphs.
Therefore, we instead batch-process multiple nodes in parallel. 
For the graph in Figure \ref{fig:sparse-graph}, we can batch-process $1$ equalizer \texttt{e}, $3$ noisegates \texttt{n}, and $5$ gain/pannings \texttt{g} sequentially.
Then, we aggregate the intermediate outputs to $2$ subgroup mixes \texttt{m} (also in parallel).
This part is identical to graph neural networks' ``message passing,'' so we adopt their implementations \cite{fey2019fast}.
We repeat these parallel computations until we reach the output node \texttt{o}. 
By doing so, we obtain the output faster; in this example, the number of sequential processing is reduced from $15$ (one-by-one) to $8$ (optimal). 
We empirically found that up to $5.8\times$ speedup can be achieved for the pruned graphs with a \texttt{RTX3090} GPU.
Refer to our companion paper for further details \cite{lee2024grafx}.

\subsection{Mixing Console}
We construct a mixing console $G_\mathrm{c}$ as follows 
(see Figure \ref{fig:full-graph}). 
\begin{enumerate}[label=(\roman*), leftmargin=7mm]
  \setlength\itemsep{0em}
  \item We add an input node \texttt{i} for each source track. 
  \item We connect a serial chain (with a fixed order) of an equalizer \texttt{e}, compressor \texttt{c}, noisegate \texttt{n}, stereo imager \texttt{s}, gain/panning \texttt{g}, multitap delay \texttt{d}, and reverb \texttt{r} for each input. 
  \item We subgroup and sum the processed tracks with mix nodes \texttt{m} based on the prepared subgrouping information.
  \item We apply the same chain \texttt{ecnsgdr} to each mix output, then pass it to the output node \texttt{o} (we omit the mix module here).
\end{enumerate}
\vspace{-1mm}

\subsection{Optimization}
Before exploring the pruning of each mixing console, as a sanity check, we first evaluate its match performance.
To investigate how much each processor type contributes to the match quality,
we start with a base graph, a mixing console with no processors that simply sums all the inputs. Then, we add each processor type one by one to the processor chain (see the first column of Table \ref{table:mixing-console}). We optimize and evaluate all these preliminary graphs for each song.
For each graph, we train its parameters and weights simultaneously with an audio-domain loss given as follows,
\begin{equation}
    L_\mathrm{a} = \alpha_{\mathrm{lr}}L_\mathrm{lr} + \alpha_{\mathrm{m}}L_\mathrm{m} + \alpha_{\mathrm{s}}L_\mathrm{s} 
\end{equation}
where each term $L_\mathrm{x}$ is a variant of multi-resolution STFT loss \cite{yamamoto2020parallel} ($\mathrm{x \in \{lr, m, s\}}$, $\mathrm{lr}$: left/right, $\mathrm{m}$: mid, $\mathrm{s}$: side)
\begin{equation}
    L_\mathrm{x} = \sum_{i=1}^I \Bigg[ 
		\frac{\| \log Y^{(i)}_\mathrm{x} - \log \hat{Y}^{(i)}_\mathrm{x} \|_1}{N} 
	    + \frac{\|Y^{(i)}_\mathrm{x} - \hat{Y}^{(i)}_\mathrm{x} \|_F}{\|Y^{(i)}_\mathrm{x}\|_F} 
		\Bigg].
\end{equation}
Here, $Y_\mathrm{x}^{(i)}$ and $\hat{Y}_\mathrm{x}^{(i)}$ denote the $i^\mathrm{th}$ Mel spectrograms of the target and predicted mixture, respectively. 
$N$, $\|\cdot\|_1$, and $\|\cdot\|_F$ denote the number of frames, $l_1$ norm and Frobenius norm, respectively.
We use FFT sizes of $512$, $1024$, and $4096$, and hop sizes are $1/4$ of their respective FFT sizes. 
The number of Mel filterbanks is set to $96$ for all scales.
We apply A-weighting before each STFT \cite{wright2020perceptual}. 
The per-channel loss weights are set to 
$\alpha_\mathrm{lr} = 0.5$, $\alpha_\mathrm{m}=0.25$, and $\alpha_\mathrm{s}=0.25$. 
The implementation is based on \texttt{auraloss} \cite{steinmetz2020auraloss}.
We further add a regularization that promotes gain-staging, 
a common practice of audio engineers that keeps the total energy of input and output roughly the same.
This is achieved with the following loss:
\begin{equation}
    L_\mathrm{g} = \sum\nolimits_{v_i \in V_\mathrm{g}} \left| \log \| f_i(u_{i})_\mathrm{m} \|_2 - \log \| u_{i, \mathrm{m}} \|_2 \right|
\end{equation}
where $(\cdot)_\mathrm{m}$ and $\|\cdot\|_2$ denote mid channel and $l_2$ norm, respectively.
We apply this regularization to a subset of processors $V_\mathrm{g} \subset V$ that comprises all equalizers, reverbs, and multitap delays.
This allows us to (i) eliminate redundant gains that these linear-time invariant (LTI) processors could create
and (ii) restrict the parameters to be in a reasonable range. 
Therefore, the total loss is given as
\begin{equation}
    L(\mathbf{P}, \mathbf{w}) = L_\mathrm{a}(\mathbf{P}, \mathbf{w}) + \alpha_\mathrm{g} L_\mathrm{g}(\mathbf{P})
\end{equation}
where the gain-staging weight is set to $\alpha_\mathrm{g} = 10^{-3}$. Here, we used a slightly different notation from Equation \ref{eq:objective} to emphasize what is optimized. 
Each console is optimized for $12\si{k}$ steps using AdamW \cite{loshchilov2017decoupled} 
with a $0.01$ learning rate.
For each step, we random-sample a $3.8\si{s}$ region of dry sources $\bfS$ (thus the batch size is $1$), compute the mix $\hat{y}$, 
and compare its last $2.8\si{s}$ with the corresponding ground-truth $y$.
Note that the first second is used only for the ``warm-up" of the processors with long states 
such as compressors and reverbs. 

\begin{table}
\setlength\tabcolsep{2.8pt}
\renewcommand{\arraystretch}{.8}
\begin{center}
\caption{
\it
Matching performances of the mixing consoles using different processor type configurations. 
}
\vspace{-.5mm}
\begin{tabular}{lc|cccc}
\toprule
 & & $L_\mathrm{a}$ & $L_\mathrm{lr}$ & $L_\mathrm{m}$ & $L_\mathrm{s}$ 
\\ \midrule
\multicolumn{2}{l|}{Base graph (sum of dry sources)}
& $19.7$ & $1.52$ & $1.46$ & $74.3$ \\
\midrule
+ Gain/panning 
& \texttt{\textcolor{white}{ecns\textcolor{black}{g}dr}}
& $.689$ & $.686$ & $.634$ & $.752$ \\
+ Stereo imager 
& \texttt{\textcolor{white}{ecn\textcolor{black}{sg}dr}}
& $.676$ & $.671$ & $.623$ & $.734$ \\
+ Equalizer 
& \texttt{\textcolor{white}{\textcolor{black}{e}cn\textcolor{black}{sg}dr}}
& $.557$ & $.549$ & $.493$ & $.637$ \\
+ Reverb 
& \texttt{\textcolor{white}{\textcolor{black}{e}cn\textcolor{black}{sg}d\textcolor{black}{r}}}
& $.481$ & $.471$ & $.457$ & $.523$ \\
+ Compressor 
& \texttt{\textcolor{white}{\textcolor{black}{ec}n\textcolor{black}{sg}d\textcolor{black}{r}}}
& $.423$ & $.407$ & $.385$ & $.492$ \\
+ Noisegate 
& \texttt{\textcolor{white}{\textcolor{black}{ecnsg}d\textcolor{black}{r}}}
& $.414$ & $.398$ & $.375$ & $.485$ \\
\midrule
+ Multitap delay (full) 
& \texttt{\textcolor{white}{\textcolor{black}{ecnsgdr}}}
& $.409$ & $.395$ & $.375$ & $.469$ \\
\bottomrule
\end{tabular}

\label{table:mixing-console}
\end{center}
\vspace{-5mm}
\end{table}

\subsection{Results}\label{section:console-results}
Table \ref{table:mixing-console} reports the evaluation results that are calculated over the entire song.
First, the base graph results in an audio loss $L_\mathrm{a}$ of $19.7$. 
The side-channel loss $L_\mathrm{s}$ is especially large as most source tracks are close to mono while the target mixes have wide stereo images. 
With the gain/pannings and stereo imagers, 
we can achieve ``rough mixes'' with a loss of $0.676$.
Then, we fill in the missing details with the remaining processor types.
Every type improves the match, and the full mixing console reports a loss of $0.409$.
Also, the top $5$ rows of Figure \ref{fig:spec-main} show mid/side log-magnitude STFTs of the target mixes, matches of the mixing consoles, and their errors. 
We report the results with $3$, $4$, $6$, and $7$ types where the choice of processors and their order follow Table \ref{table:mixing-console};
see the supplementary page for the results on other configurations and additional songs.
Again, we can observe that adding each type improves the match from the spectrogram error plots.
Furthermore, each song benefits more from different types; for the song \texttt{RockSteady}, 
the multitap delays improve the match more than the reverbs (Figure \ref{fig:spec-b}), which is different from the average trend. 
Yet, this is expected since the original mix heavily uses the delay effects.
Finally, we note that mixes from \texttt{MixingSecrets} are more challenging to match than the others; it reports a mean audio loss of $0.545$, while \texttt{MedleyDB} and \texttt{Internal} report $0.296$ and $0.385$, respectively. 

\vspace{-4mm}
\section{Music Mixing Graph Search}
\vspace{-.5mm}
Considering the full mixing console $G_\mathrm{c}$ as an upper bound in terms of the matching performance, 
we want to find a sparser graph with a similar match quality.
We achieve this by pruning the console as much as possible
while keeping the loss increase up to a tolerance threshold $\tau$. 
This objective can be written as 
\begin{equation}\label{eq:pruning}
    \mathrm{minimize} \;\; | V_\mathrm{p} |  \quad \mathrm{s.t.} \;\; 
	\min L_\mathrm{a} (G_\mathrm{p}) \leq \min L_\mathrm{a}(G_\mathrm{c}) + \tau
\end{equation}
where $V_\mathrm{p}$ and $|\cdot|$ denote the pruned graph's node set and its cardinality, respectively.
We define the pruning as removal of the nodes $V_\mathrm{c}\setminus V_\mathrm{p}$ 
and re-routing of their edges,
in a way that is equivalent to setting them to ``bypass,'' i.e., $w_i=0$ for  $v_i \in V_\mathrm{c} \setminus \Vp$.
Also, $\min (\cdot)$ signifies that we are (ideally) interested in the optimized audio loss. 
We only prune the processors, not the auxiliary nodes. 
Hence, we define a pruning ratio $\rho$ as the number of pruned processors divided by the number of processors in the initial console. 

\subsection{Iterative Pruning}
Finding the optimal (sparsest) solution $\Vp^*$ is prohibitively expensive.
First, due to the interaction between the processors, we need a combinatorial search.
As such, we instead assume their independence and prune the processors in a greedy manner. 
Following the iterative approach \cite{castellano1997iterative},
we gradually remove processors whenever the tolerance condition is satisfied.
Under this setup, we still need to fine-tune intermediate pruned graphs before evaluating the tolerance condition.
For reasonable computational complexity, we simply omit this fine-tuning, paying the cost of possibly missing more removable processors. 
Our method is summarized in Algorithm \ref{algo:pruning} (in the following parentheses denote line numbers).
First, we construct a mixing console $\Gc = (V, E)$, optimize its parameters $\bfP$ and dry/wet weights $\bfw$, and evaluate the loss (\ref{algo:prune:init}-\ref{algo:prune:console-eval}). 
This validation loss $\La^\mathrm{min}$ serves as a pruning threshold with the tolerance $\tau$. 
Then, we alternate between pruning and fine-tuning, i.e., further optimization of the remaining parameters and weights
(\ref{algo:prune:iter-start}-\ref{algo:prune:iter-end}). 
Each pruning stage consists of multiple trials, 
which sample subsets of candidates ${V}_\mathrm{cand}$ from the set of remaining processors $V_\mathrm{pool}$ (\ref{algo:prune:sample}) and check whether they are removable (\ref{algo:prune:trial-if}).
We keep the pruning if the result satisfies the constraint or cancel it otherwise 
(\ref{algo:prune:trial-if}-\ref{algo:prune:trial-endif}).
We repeat this process until the terminal condition (\ref{algo:prune:while}) is satisfied.
Implementation-wise, we multiply binary masks, $\bfm$ and ${\bfm}_{\:\!\mathrm{cand}}$, to the weight vector $\bfw$
to mimic the pruning during the trials (\ref{algo:prune:try}). 
After that, we actually update the graph and remove the pruned processors' parameters and weights
for faster search (\ref{algo:prune:actual-prune}).
Sometimes, albeit rare, the pruning can improve the match. 
In this case, we update the threshold (\ref{algo:prune:update}).

\begin{algorithm}[t]
\caption{
\it
Music mixing graph search with iterative pruning.
}
\begin{algorithmic}[1]
\Require A mixing console $\Gc$, dry tracks $\bfS$, and mixture $y$
\Ensure Pruned graph $\Gp$, parameters $\bfP$, and weights $\bfw$
\vspace{.7mm}

\State $\bfP, \bfw \gets \fn{Initialize}(\Gc)$ 
\label{algo:prune:init}

\State $\bfP, \bfw \gets \fn{Train}(\Gc, \bfP, \bfw, \bfS, y)$

\State $\La^\mathrm{min} \gets \fn{Evaluate}(\Gc, \bfP, \bfw, \bfS, y)$ 
\label{algo:prune:console-eval}

\State $\Gp \gets \Gc$

\For{$n$ $\leftarrow$ $1$ to $N_\mathrm{iter}$} \label{algo:prune:iter-start} 
	\State $V_\mathrm{pool}, \bfm \gets \fn{GetAllProcessors}(V), \mathbf{1}$
	\While {$\mathrm{TryPrune}\:\!(V_\mathrm{pool}, \bfw, \bfm)$} \label{algo:prune:while} 
		\State $V_\mathrm{cand}, {\bfm}_{\:\!\mathrm{cand}} \gets \fn{SampleCandidate}(V_\mathrm{pool}, \bfw)$ \label{algo:prune:sample}
		\State $\La \gets \fn{Evaluate}(\Gp, \bfP, \bfw\odot \bfm \odot {\bfm}_{\:\!\mathrm{cand}}, \bfS, y)$ \label{algo:prune:try}
		\If {$\La < \La^\mathrm{min} + \tau$} \label{algo:prune:trial-if}
			\State $\La^\mathrm{min} \gets \min (\La^\mathrm{min}, \La) $ \label{algo:prune:update}
                \State $\bfm \gets \bfm \odot {\bfm}_{\:\!\mathrm{cand}}$
		\EndIf \label{algo:prune:trial-endif}
		\State $V_\mathrm{pool} = \fn{UpdatePool}(V_\mathrm{pool}, V_\mathrm{cand})$ \label{algo:prune:update-pool}
    \EndWhile
	\State $\Gp, \bfP, \bfw \gets \fn{Prune}(\Gp, \bfP, \bfw, \bfm)$ \label{algo:prune:actual-prune}
    \State $\bfP, \bfw \gets \fn{Train}(\Gp, \bfP, \bfw, \bfS, y)$
\EndFor \label{algo:prune:iter-end}
\State \Return $\Gp, \bfP, \bfw$
\end{algorithmic}
\label{algo:pruning} 
\end{algorithm}

\subsection{Candidate Sampling}\label{section:candidate}
The remaining design choices are choosing an appropriate candidate set $\Vc$ 
(\ref{algo:prune:sample}, \ref{algo:prune:update-pool}) and deciding when to terminate the trials (\ref{algo:prune:while}).
We explore the following $3$ approaches.
\begin{itemize}[leftmargin=4mm]
  \setlength\itemsep{0em}
    \item \textbf{Brute-force ---} 
		We random-sample every processor one by one, i.e., $|\Vc|=1$.
		This granularity could achieve high sparsity, but comes with a large computational cost.

    \item \textbf{Dry/wet ---} 
		For efficient pruning, we need an informed guess of each node's importance.
		Intuitively, we can use each dry/wet weight $w_i$ as an approximate importance.
		This observation leads to the following. For each pruning iteration:
		\begin{enumerate}[label=(\roman*), leftmargin=7mm]
		  \setlength\itemsep{0em}
			\item 
                    We create a set of remaining processor types $\Tc$.
                    Next, we count the number of processors of each type $t \in \Tc$, denoted as $N_t$.
		    \item For each trial, we sample a type $t \in \Tc$
				and choose the smallest-weight processors of that type as candidates.                    
				The number of candidates is set to $|\Vc| = \max\;\!(1, \lfloor r_t N_t \rceil)$ where $r_t$ denotes the portion of the chosen processors and is initialized to $0.1$ for every pruning iteration. \label{step-2}
    
			\item When the trial fails, we perform one of the following. If $|\Vc| > 1$,
				we halve the candidate set, i.e., $r_t \gets r_t/2$.
				Otherwise, i.e., if $|\Vc| = 1$, we finish the search of this type by removing it from the pool as  $\Tc\gets\Tc\setminus\{t\}$.\label{step-3}
                
			\item We iterate above two \ref{step-2}-\ref{step-3} until $\Tc=\emptyset$.
		\end{enumerate}
		This way, we can skip large-weight nodes and evaluate multiple candidates, reducing the total number of trials. 
            Note that if we set $r_t=0.5$, this method is similar to the simple binary search. However, it can lead to over-pruning of specific types sampled early in \ref{step-2}. Hence, we set $r_t$ to a more conservative value $0.1$.

    \item \textbf{Hybrid ---} 
		Solely relying on the weight values could miss some processors 
		that can be pruned but have large weights.
		We mitigate this by combining the above two, 
		running the brute-force method for every $4^\mathrm{th}$ iteration.
\end{itemize}
By default, we use the hybrid method with tolerance $\tau = 0.01$.

\begin{figure*}[t]
    \begin{center}{\includegraphics[width=2.08\columnwidth]{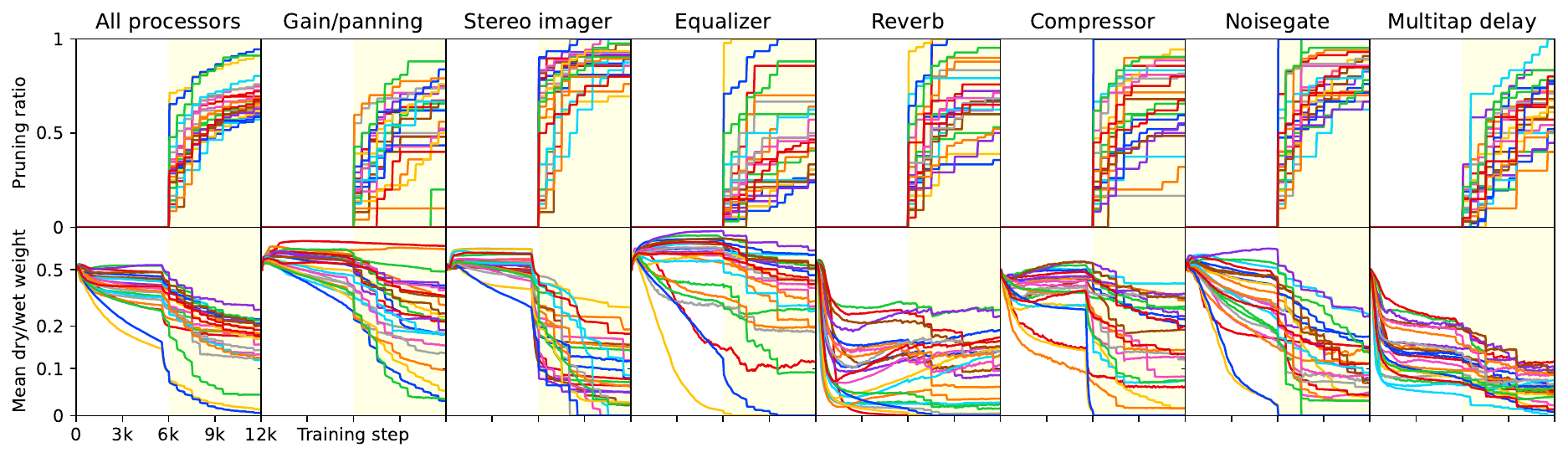}}
    \vspace{-2.2mm}
        \caption{
        \it
			Process of iterative pruning
			(hybrid, $\tau=0.01$). 
			  $24$ songs ($8$ songs per dataset) are shown; each color represents an individual song. 
			The upper and lower rows show the pruning ratios
			and mean dry/wet weights. 
			The yellow-shaded regions show the pruning phase.
        } 
        \label{fig:pruning-process}
    \end{center}
    \vspace{-5.2mm}
\end{figure*}

\subsection{Optimization}
We use identical audio loss $\La$ and gain-staging regularization $L_\mathrm{g}$.
To promote sparsity, we add a weight regularization $L_\mathrm{p}$, 
a $l_1$ norm of the weight $\mathbf{w}$. 
Hence, the full objective is as follows,
    \begin{equation}
        L(\bfP, \bfw) = \La(\bfP, \bfw) + \alpha_\mathrm{g} \Lg(\bfP) + \alpha_\mathrm{p} \Lp(\bfw).
    \end{equation}
We first train the console with $6\si{k}$ steps. 
Then, we repeat $N_\mathrm{iter}=12$ rounds of pruning, each with $0.5\si{k}$-step fine-tuning. 
As a result, the total number of optimization steps is the same as the previous console training.
During the first $4\si{k}$ steps of the pruning phase, 
we linearly increase the sparsity coefficient $\alpha_\mathrm{p}$ from $0$ to $10^{-4}$. 
While we halved the full console optimization steps, which could lead to increased loss,
it is justified due to the tight resource constraints.
With a \texttt{RTX3090} GPU, each song took about $56\si{m}$, $29\si{m}$, and $36\si{m}$ using 
the brute-force, dry/wet, and hybrid methods, respectively.

\subsection{Results} \label{section:pruning-results}
\paragraph{Pruning process} 
Figure \ref{fig:pruning-process} shows how the pruning progresses.
Each graph's sparsity increases gradually while its weights adapt over time.
This trend is different for different processor types.
The mean objective metrics are reported in Table \ref{table:pruning}.
The default setting reports an average audio loss $\La$ of $0.422$, an $0.013$ increase from the full consoles, 
slightly exceeding the tolerance $\tau =0.01$.
This was expected due to the shorter full console training.
The average pruning ratio $\rho$ is $0.67$
and the equalizer and stereo imager are the most and least remaining types 
($0.46$ and $0.86$), respectively.
We note that \texttt{MedleyDB} and \texttt{MixingSecrets} report similar pruning ratios of $0.61$ and $0.62$, respectively. However, the \texttt{Internal} graphs are more sparse; their average pruning ratio is $0.77$.


\begin{table}[t]
\setlength\tabcolsep{2.8pt}
\renewcommand{\arraystretch}{.8}
\begin{center}
\caption{\it Pruning results with various candidate selection methods and tolerance $\tau$. 
The subscripts denote per-type pruning ratios.
}
\begin{tabular}{ll|ccccccccc}
\toprule
& \multicolumn{1}{c|}{$\tau$} & $L_\mathrm{a}$
& $\rho$ 
& $\rho_\textttt{g}$ 
& $\rho_\textttt{s}$ 
& $\rho_\textttt{e}$ 
& $\rho_\textttt{r}$ 
& $\rho_\textttt{c}$ 
& $\rho_\textttt{n}$ 
& $\rho_\textttt{d}$ 
\\

\midrule
Mix console
& \multicolumn{1}{c|}{$-$}
& $.409$ 
& $-$ 
& $-$ & $-$ & $-$ & $-$ & $-$ & $-$ & $-$ 
\\

\midrule
Brute-force & $.01$ 
& $.424$ 
& $.69$ 
& $.54$ & $.85$ & $.53$ & $.76$ & $.71$ & $.78$ & $.69$ 
\\

\midrule
{Dry/wet} & $.01$ 
& $.420$ 
& $.62$ 
& $.51$ & $.84$ & $.38$ & $.69$ & $.66$ & $.76$ & $.53$ 
\\

\midrule
\multirow{3}{*}{Hybrid} & $.001$ 
& $.411$
& $.49$ 
& $.35$ & $.76$ & $.27$ & $.53$ & $.57$ & $.62$ & $.34$ 
\\

& $.01$   
& $.422$ 
& $.67$  
& $.51$ & $.86$ & $.46$ & $.71$ & $.71$ & $.79$ & $.63$ 
\\

& $.1$  
& $.499$  
& $.87$  
& $.73$ & $.94$ & $.81$ & $.90$ & $.85$ & $.91$ & $.92$ 
\\

\bottomrule
\end{tabular}
\vspace{-5.3mm}
\label{table:pruning}
\end{center}
\end{table}

\begin{figure}[!t]
  \centering
  \vspace{-1.4mm}
  \includegraphics[width=.98\columnwidth]{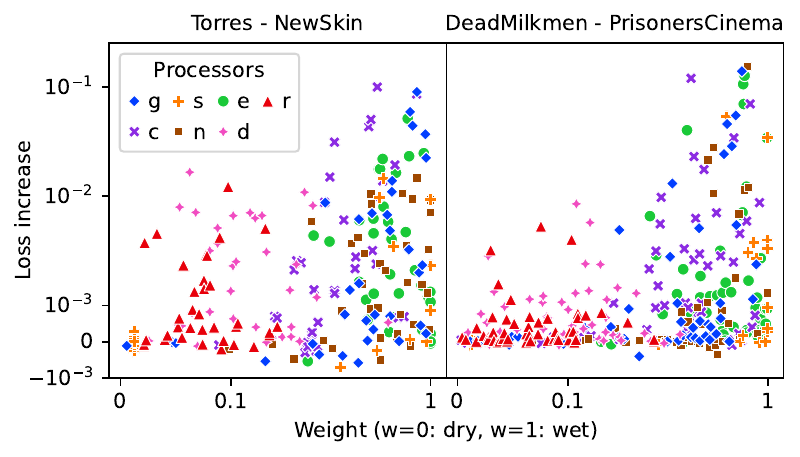}
  \vspace{-2.5mm}
  \caption{
  \it
	  Each node's weight and loss increase when pruned.
  }
  \label{fig:loss-vs-weight} 
  \vspace{-1.6mm}
\end{figure}

\begin{figure*}
    \begin{center}
    \captionsetup[subfloat]{captionskip=8pt}
    \subfloat[$\tau=0.001$\\$\rho = 0.45, \La=0.525$ \label{fig:pruned-0.001}]{
        \hspace{3.8mm}
        \includegraphics[height=.535\columnwidth]{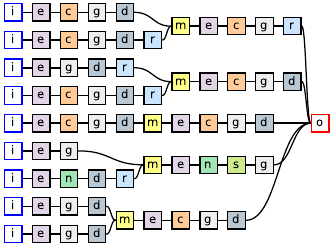}
        \hspace{3.8mm}
        }
    \subfloat[$\tau=0.01$\\$\rho = 0.74, \La=0.525$ \label{fig:pruned-0.01}]{
        \includegraphics[height=.535\columnwidth]{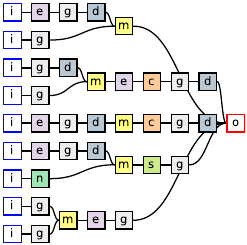}
        }
    \subfloat[$\tau=0.1$\\$\rho = 0.85, \La=0.611$
    \label{fig:pruned-0.1}]{
        \hspace{3.8mm}
        \includegraphics[height=.535\columnwidth]{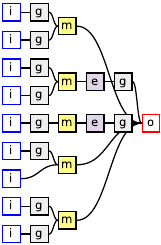} 
        \hspace{3.8mm}
        } 
    \vspace{-1.5mm}
    \caption{
    \it
		Pruning results (hybrid method) with various tolerances. 
		Song: \texttt{TablaBreakbeatScience\_RockSteady}.
	}
    \label{fig:pruned-graphs} 
    \end{center}
    \vspace{-6mm}
\end{figure*}

\paragraph{Sampling method comparison} 
Here, we fix the tolerance $\tau$ to $0.01$ and compare the candidate sampling approaches; see Table \ref{table:pruning}.
As expected, the brute-force method achieves the highest sparsity, 
reporting a pruning ratio of $0.69$. 
Its average audio loss is also the highest, $0.424$, an $0.015$ increase from the mixing console result.
The dry/wet method prunes the least with $0.62$, 
$7\%$ lower than the brute-force method.
However, its audio loss is the lowest, $0.420$,
as more processors remained.
We can investigate the cause of this difference in sparsity by analyzing the relationship between each dry/wet weight $w_i$ and the loss increase $\Delta_i$ caused by pruning the processor $v_i$ defined as follows,
\begin{equation}
    \Delta_i = \La(G\setminus\{v_i\}) - \La(G).
\end{equation}
Figure \ref{fig:loss-vs-weight} shows scatterplots for $2$ random-sampled songs, one for each song. 
Each point $(w_i, \Delta_i)$ corresponds to each processor after the initial console training.
To maximize the sparsity using the dry/wet method, a monotonic relationship between the weights $w_i$ and loss increases $\Delta_i$ is desirable, which is unfortunately not the case.
Yet, a positive correlation exists, and this becomes more evident when we analyze the relationship for each type separately, justifying the per-type candidate selection. 
Still, we cannot completely remove the weakness of the dry/wet method,
leading us to the hybrid strategy as a compromise.
We note that the pruning methods are not only different in sparsity but also in trade-offs between sparsity and match performance. 
By evaluating the methods with more fine-grained tolerance settings ($7$ values from $0.001$ to $0.2$), we observed that the brute-force method finds graphs with better matches even with the same graph size, closely followed by the hybrid method; refer to the supplementary page for the details.


\paragraph{Choice of tolerance}
Finally, we analyze the effect of the value of tolerance  $\tau$. 
Even with a very low tolerance $\tau=0.001$, we can nearly halve the number of processors, i.e., $\rho = 0.49$.
If we set the value too high, e.g., $\tau =0.1$, the resulting graphs are highly sparse 
but degrade their matches ($\La=0.499$, i.e., $0.090$ increase).
The default setting of $\tau=0.01$ seems ``just right,'' balancing the match performance and graph sparsity.
We can verify this with the spectrogram errors (bottom $3$ rows for each subplot; see Figure \ref{fig:spec-main} and supplementary page). 
There is no noticeable degradation from the full consoles to $\tau = 0.001$ and $0.01$.


\vspace{1.5mm}
\paragraph{Case study}
We report the pruning method's behavior from observations of the individual results.

\begin{itemize}[leftmargin=4mm]
\setlength\itemsep{0em}
        \item 		
            Recall that, for the song \texttt{RockSteady},
		there was no clear performance improvement when we added the reverbs (Figure \ref{fig:spec-b}).
		Hence, we can expect those reverbs to be pruned with a moderate tolerance $\tau$.
		Figure \ref{fig:pruned-graphs} shows that this is indeed the case;
		only $5/14$ reverbs are left when $\tau=0.001$ and $0/14$ for $\tau = 0.01$,
		which is much less than the average statistics
        (Table \ref{table:pruning} and \ref{table:pruning-main-per-dataset}).
	When $\tau=0.1$, processors for the details get removed; 
        only the gain/pannings and equalizers remain.
            See captions in Figure \ref{fig:pruned-graphs} for the pruning ratios and audio losses of the pruned graphs (the full console has an audio loss of $0.523$).

	\item 
            The current pruning method fails to detect some redundant processors.
		In Figure \ref{fig:pruned-0.01}, the bottom $2$ sources are processed with $3$ gain/pannings. 
		Since there is no nonlinear or time-varying processor between those,
		at least one can be ``absorbed'' by the others.
		While this case can be handled with some post-processing, 
            it hints that we might have missed more sparse graphs.
            
	\item 
		Each pruning of the same song yields a slightly different graph. 
            Pruning a mixing console of \texttt{GirlOnABridge} multiple times resulted in graphs with the number of processors from $19$ to $22$.
  		This is because our iterative pruning has a stochastic and greedy nature; 
            candidates that were sampled early are more likely to be pruned. 
            Refer to the supplementary page for the pruned graphs.

	\item
		The pruning does not necessarily result in 
		graphs that are close to the maximum loss $\La(\Gc)+\tau$.
            For \texttt{RockSteady}, pruning with $\tau=0.01$ resulted in a loss of $0.525$, much lower than the threshold.
            Interestingly, the $\tau=0.001$ case achieved the same loss in spite of a much lower pruning ratio ($0.56$ versus $0.74$).

	\item 
		Processors for sources with short spans and low energy tend to get pruned 
            as their contributions to the audio loss are small.
		Yet, we found that this could sometimes be perceptually noticeable.

\end{itemize}
\begin{table}[t]
\vspace{1.5mm}
\setlength\tabcolsep{2.74pt}
\renewcommand{\arraystretch}{.8}
\begin{center}
\vspace{-.5mm}
\caption{
\it
Pruning results with the default setting on the full dataset.
}
\vspace{-.5mm}
\begin{tabular}{l|ccccccccc}
\toprule
& $L_\mathrm{a}$
& $\rho$ 
& $\rho_\textttt{g}$ 
& $\rho_\textttt{s}$ 
& $\rho_\textttt{e}$ 
& $\rho_\textttt{r}$ 
& $\rho_\textttt{c}$ 
& $\rho_\textttt{n}$ 
& $\rho_\textttt{d}$ 
\\

\midrule
\texttt{MedleyDB}
& $.431$ 
& $.63$ 
& $.37$ & $.84$ & $.44$ & $.69$ & $.74$ & $.77$ & $.57$ 
\\

\texttt{MixingSecrets}
& $.625$ 
& $.64$ 
& $.50$ & $.87$ & $.33$ & $.64$ & $.63$ & $.80$ & $.69$ 
\\

\texttt{Internal}
& $.434$ 
& $.75$ 
& $.70$ & $.87$ & $.55$ & $.73$ & $.85$ & $.86$ & $.72$ 
\\
\midrule

\texttt{Total}
& $.506$ 
& $.69$ 
& $.57$ & $.87$ & $.45$ & $.69$ & $.75$ & $.82$ & $.69$ 
\\

\bottomrule
\end{tabular}
\vspace{-2.1mm}
\label{table:pruning-main-per-dataset}
\end{center}
\end{table}

\begin{figure}[t]
  \centering
  \vspace{-1.6mm}
  \includegraphics[width=.91\columnwidth]{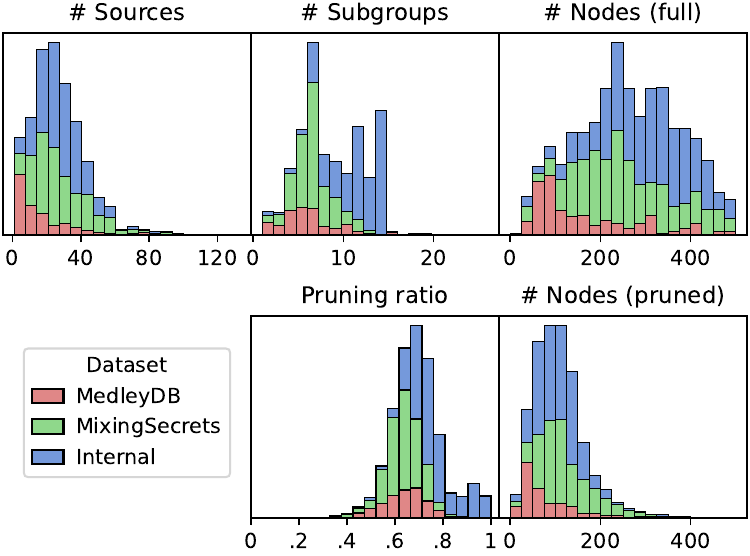}
  \vspace{-1.3mm}
  \caption{
  \it
	  Statistics of the consoles and pruned graphs (full data).
	  Each dataset's results are stacked to form the full histograms.
  }
  \label{fig:pruned-stats} 
  \vspace{-3mm}
\end{figure}

\paragraph{Full results} \label{section:full-pruning}
Finally, we pruned every song in the full dataset ensemble.  
Table \ref{table:pruning-main-per-dataset} reports the results.
The overall trend follows the evaluation subset results 
but with a higher mean audio loss ($0.509$ compared to the previous $0.422$).
Figure \ref{fig:pruned-stats} shows statistics of the $3$ datasets, initial mixing console graphs, and their pruned versions.
\texttt{MedleyDB} has the smallest number of source tracks, an average of $17.6$.
The \texttt{Internal} dataset has the largest ($28.8$), closely followed by the \texttt{MixingSecrets} ($27.9$).
The \texttt{Internal} also has more subgroups, resulting in even larger mixing consoles.
This is one potential cause of the higher sparsity of its pruned graphs; more processors were initially used to match the mix, and many of them were redundant.
On average, $72.1$ processors ($108.5$ nodes) were remained for each song. Since each full mixing console has an average of  $247.6$ processors ($280.1$ nodes), we achieved a pruning ratio of $0.692$.

\section{Discussion}
\paragraph{Summary}
We started with a general formulation of retrieving mixing graphs from dry sources and mix. 
Then, we posed restrictions to restrict the search to the pruning of mixing consoles, making it computationally feasible and obtaining more interpretable graphs. 
Next, with additional assumptions, we derived the iterative method that gradually removes negligible processors in a stochastic and greedy manner. 
As a result, instead of finding the exact optimal, our method gives (or ``samples'') one of the close-to-optimal graphs. 
With the differentiable processors and relaxation of the pruning with the dry/wet weights, 
we optimized this objective via gradient descent.
We explored $3$ methods to choose pruning candidates, comparing them in terms of their computational cost and graph sparsity.
The hybrid method gave a good compromise, so we used it to gather over one thousand graph-audio pairs.

\vspace{2.7mm}
\paragraph{Future works} 
We list possible extensions of our method.
\begin{itemize}[leftmargin=4mm]
\setlength\itemsep{0em}
\item The choice of processors and their implementations directly affect the match quality.
Our setup, 
including the equalizer with zero-phase FIR and the reverb based on STFT mask, 
was motivated by its simplicity and fast computation on a GPU.
However, there are other options, such as parametric equalizer \cite{nercessian2020neural} and artificial reverberation \cite{lee2022differentiable}, that allow more efficient computation in the CPU and have compact parameterizations.
Also, the spectrogram errors showed clear temporal patterns (vertical stripes), indicating that the loudness dynamics were not precisely matched. 
We suspect it is due to the ballistics approximation error, as recently reported \cite{steinmetz2023high}. If so, we might need a more sophisticated implementation of the compressor and noisegate.
In addition, the current method did not consider time-varying parameters (or ``automation''), which can cause audible errors.
For example, we could not match fade-out, i.e., a gradual decrease in track loudness.
Finally, we can add other processor types, e.g., saturation/distortion \cite{colonel2022reverse} or modulation effects \cite{carson2023differentiable}. 

\item
We note several considerations to improve the current pruning method in terms of sparsity, match quality, and interpretability.
First, we can modify the mixing console to
reflect real-world practices more.
For example, we can add send and return loops with additional processor chains. 
Post-equalizers for compressors and processors with multiple inputs or outputs (e.g., auxiliary sidechain and crossover filter) are also commonly used.
Second, to prevent the pruning from harming the perceptual quality, the tolerance condition and the objective function must be appropriately designed. We used a simple multi-resolution STFT loss \cite{yamamoto2020parallel, steinmetz2020auraloss}, which has been reported to miss some perceptual features 
\cite{hayes2023review, turian2020m}.
Therefore, we might need an alternative objective as a remedy \cite{vahidi2023mesostructures}.
Third, as discussed before, using average loss to determine the pruning might be inappropriate.
Lastly, to increase the sparsity, more advanced neural network pruning techniques \cite{he2023structured, cheng2023survey}
and domain-specific post-processing, e.g., merging LTI processors to a single processor with the combined effect, can be applied.

\item 
We may relax the prior assumptions and restrictions on graph structures. This will expand our search space and require different search methods other than pruning. For example, allowing arbitrary processor order extends our framework to different architecture search \cite{liu2018darts, ye2023fm}. A completely different approach
based on reinforcement learning could also be possible \cite{you2018graph}.
While all of these are promising, balancing flexibility, match quality, and computation cost will be the main challenge. 

\end{itemize}




\begin{figure}
  \centering
    \subfloat[\it \texttt{Torres\_NewSkin} \label{fig:spec-a}]{
        \includegraphics[clip,trim=.2mm .1mm .2mm .1mm, width=.988\columnwidth]{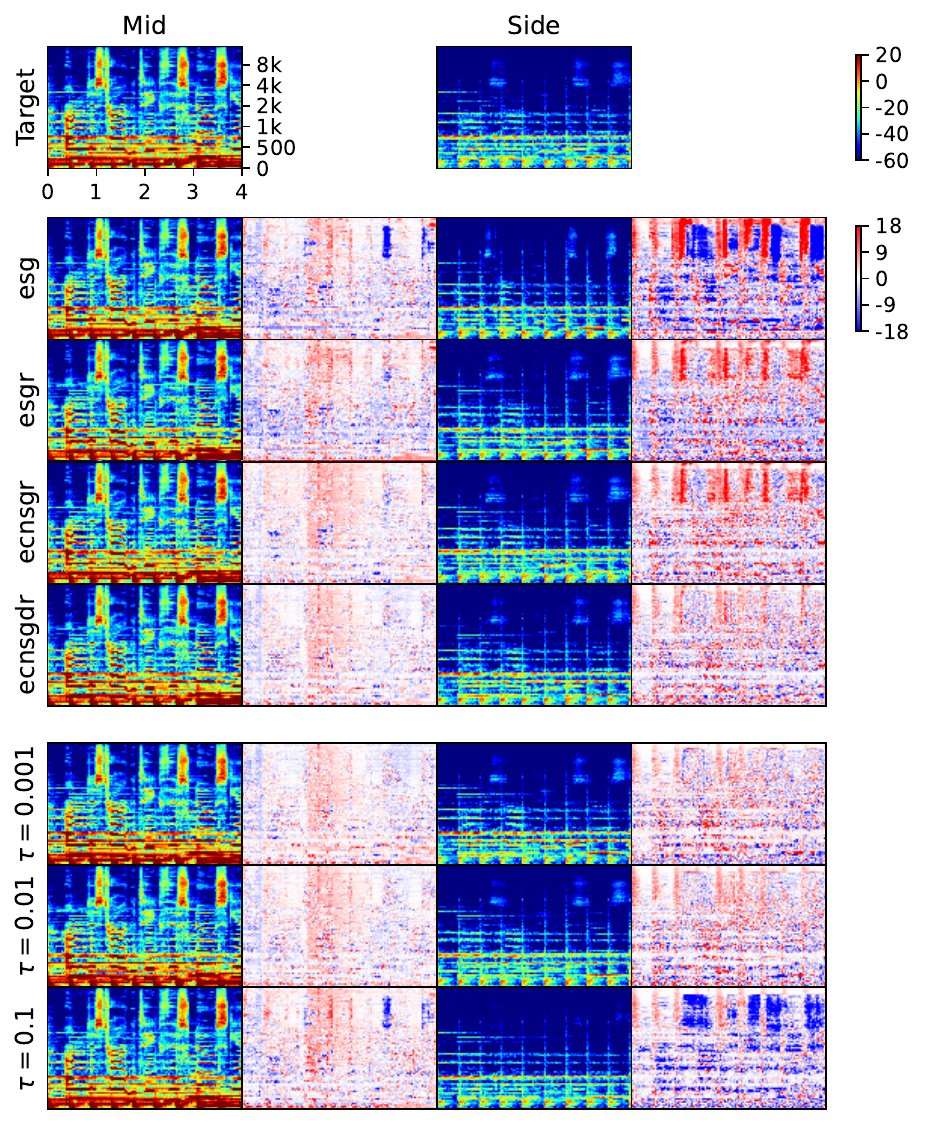}} \\
        \vspace{-2.08mm}
    \subfloat[\it \texttt{TablaBreakbeatScience\_RockSteady} \label{fig:spec-b}]{
        \includegraphics[clip,trim=.2mm .1mm .2mm .1mm, width=.988\columnwidth]{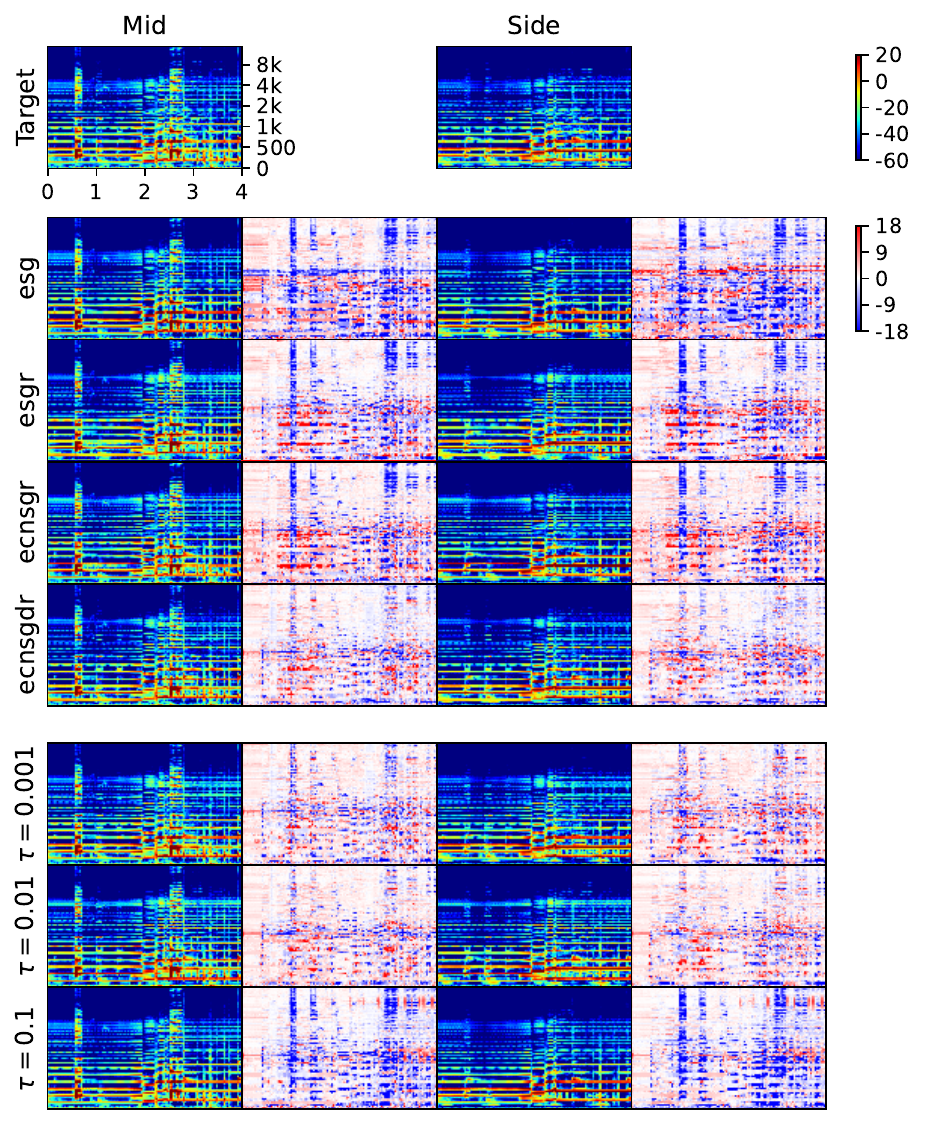}}\\
        \vspace{-0.4mm}
  \caption{
  \it
        Log-magnitude spectrograms of the matched mixes (odd columns) of mixing consoles ($4$ center rows) and pruned graphs ($3$ bottom rows; in $\si{dB}$). The even columns show the match errors.
  }
    \label{fig:spec-main} 
\end{figure}

\bibliographystyle{IEEEbib}
\bibliography{refs} 
\fi

\ifappendix
\appendix
\newpage
\section{Related Works}\label{setion:related_works}
\subsection{Composition of audio processors}\label{section:related_graph}
\looseness=-1
Most audio processors are designed to modify some specific properties of their input signals,
e.g., magnitude response, loudness dynamics, and stereo width.
As such, combining multiple processors is a common practice to achieve the full effect.
Following the main text, we will use the terminology ``graph'' to represent this composition, although some previous works considered simple structures that allow a more compact form, e.g., a sequence. 
Now, we outline the previous attempts that tried to estimate the processing graph or its parameters from reference audio. 
These works differ in task and domain, processors, graph structure, and estimation methods.
For example, if the references are dry sources and a wet mixture, this task becomes reverse engineering \cite{lee2023blind, Mitcheltree_2021, guo2023automatic, colonel2023music}.
In terms of the prediction targets, some fixed the graph and estimated only the parameters \cite{steinmetz2020diffmixconsole, uzrad2024diffmoog, colonel2023music, engel2020ddsp, ramirez2021differentiable}.
Others tried to predict the graph \cite{guo2023automatic} or both \cite{lee2023blind, Mitcheltree_2021, ye2023fm, caspe2022ddx7}.
Table \ref{table:related_works} summarizes such differences.

\subsection{Differentiable signal processing} 
\looseness=-1
\paragraph{Differentiable processor}
Exact implementation or approximation of processors in an automatic differentiation framework, e.g., \texttt{pytorch} \cite{paszke2019pytorch}, enables parameter optimization via gradient descent.
Numerous efforts have been focused on converting existing audio processors to their differentiable versions 
\cite{colonel2023music, steinmetz2022style, colonel2022reverse, carson2023differentiable, nercessian2020neural, lee2022differentiable};
refer to the recent review \cite{hayes2023review} and references therein for more details.
In many cases, these processors are combined with neural networks,
whose computation is done in GPU. 
Thus, converting the audio processors to be ``GPU-friendly''
has been an active research topic.
For example, for a linear time-invariant (LTI) system with a recurrent structure, 
we can sample its frequency response to approximate its infinite impulse response (IIR) instead of directly running the recurrence; the former is faster than the latter \cite{nercessian2020neural, lee2022differentiable}.
However, it is nontrivial to apply a similar trick to nonlinear, recurrent, or time-varying processors. 
Typically, further simplifications and approximations are employed, e.g., replacing the nonlinear recurrent part with an IIR filter \cite{steinmetz2022style} or assuming frame-wise LTI to a linear time-varying system \cite{carson2023differentiable}.
Sometimes, we can only access input and output signals.
In such a case, one can approximate the gradients with finite difference methods \cite{ramirez2021differentiable, steinmetz2023high}
or use a pre-trained auxiliary neural network that mimics the processors \cite{steinmetz2020diffmixconsole}. 
In the literature, these are also referred to as ``differentiable;'' hence, it is rather an umbrella term encompassing all methods that obtain the output signals or gradients within a reasonable amount of time.
Nevertheless, our work limits the focus to the implementations in the automatic differentiation framework.

\vspace{2mm}
\paragraph{Audio processing graph}
Now, consider a composition of multiple differentiable processors; the entire graph remains differentiable due to the chain rule.
However, the following practical considerations remain. 
If we fix the processing graph prior to the optimization and the graph size is relatively small, we can implement the ``differentiable graph'' following the existing implementations \cite{engel2020ddsp, uzrad2024diffmoog}. 
That is, we compute every processor one by one in a pre-defined topological order. 
However, we have the following additional requirements. 
First, the pruning changes the graph during the optimization. 
Therefore, our implementation must take a graph and its parameters along with the source signals as input arguments for every forward pass. 
Note that this feature is also necessary when training a neural network that predicts the parameters of any given graph \cite{lee2023blind}.
Second, the size of our graphs is much larger than the ones from previous works \cite{engel2020ddsp, uzrad2024diffmoog, ye2023fm, lee2023blind}.
In this case, the one-by-one computation severely bottlenecks the computation speed. 
Therefore, we derived a flexible and efficient graph computation algorithm (i) that can take different graphs for each forward pass as input and (ii) performs batched processing of multiple processors within a graph, utilizing the parallelism of GPUs. 
The details of this method are described in our companion paper \cite{lee2024grafx}.
Finally, we note that other than the differentiation with respect to the input signals $\bfS$ and parameters $\bfP$, one might be interested in differentiation with respect to the graph structure $G$.
The proposed pruning method performs this to a limited extent; deletion of a node $v_i$ is a binary operation that modifies the graph structure. 
We relaxed this to a continuous dry/wet weight $w_i$ and optimized it with the audio loss $\La$ and regularization $L_\mathrm{p}$.

\subsection{Graph search} \label{section:related_graph_search}
\looseness=-1
Several independent research efforts in various domains exist that search for graphs that satisfy certain requirements.
For example, neural architecture search (NAS) aims to find a neural network architecture that achieves improved performance \cite{elsken2019neural}. 
In this case, the search space consists of graphs, with each node (or edge) representing one of the primitive neural network layers.
One particularly relevant work to ours is a differentiable architecture search (DARTS) \cite{liu2018darts}, which relaxes the choice of each layer to a categorical distribution and optimizes it via gradient descent. 
Theoretically, our method can be naturally extended to this approach; we only need to change our $2$-way choice (prune or not) to $(N+1)$-way (bypass or select one of $N$ processor types).
DARTS is clearly more flexible and general, allowing an arbitrary order of processors. 
However, it also greatly increases the computational cost, as we must compute all $N$ processors to compute their weight sum for every node. 
For example, if we want to keep the mixing console structure and allow arbitrary processor choices, the memory complexity becomes $O(N^2)$ instead of the current $O(N)$.
In other words, we must pay additional costs to increase the size of the search space.
This cost increase is especially critical to us since we have to find a graph for every song.
Another popular related domain is the generation/design of molecules with desired chemical properties \cite{elton2019deep}.
One dominant approach for this task is to use reinforcement learning (RL), which estimates each graph by making a sequence of decisions, e.g., adding nodes and edges \cite{you2018graph}.
RL is an attractive choice since we can be completely free with prior assumptions on graphs, and we can use arbitrary quality measures that are not differentiable.
We also note that RL can be used for NAS \cite{zoph2016neural}.
However, applying RL to our task has a risk of obtaining nontrivial mixing graphs that are difficult for practitioners to interpret; we may need a soft regularization penalty that guides the generation process towards familiar structures, e.g., ones like the pruned mixing consoles.
Also, it may need much larger computational resources to explore the search space sufficiently.

\vspace{2mm}
\section{Dry/wet Pruning Algorithm} \label{section:dry-wet}
Algorithm \ref{algo:pruning-weight} describes the details of the dry/wet method.
For a simpler description, we modified the initialization 
to include per-type node sets and weights
as in line \ref{algo:prune-weight:init-start}-\ref{algo:prune-weight:init-end}.
The termination condition is given in line \ref{algo:prune-weight:while}.
The trial candidate sampling is implemented in line 
\ref{algo:prune-weight:sample-start}-\ref{algo:prune-weight:sample-end}.
The candidate pool update is expanded to handle the trial successes and failures separately,
shown in line \ref{algo:prune-weight:update-success} 
and \ref{algo:prune-weight:update-fail-start}-\ref{algo:prune-weight:update-fail-end}.
\clearpage
\begin{table*}
\setlength\tabcolsep{2.8pt}
\renewcommand{\arraystretch}{.8}
\begin{center}
\fontsize{8.38}{8.38}\selectfont
\caption{
\it
A brief summary and comparison of previous works on estimation of compositional audio signal processing.
}
\begin{tabular}{llp{14cm}}
\toprule
\arrayrulecolor{lightgray}
\cite{uzrad2024diffmoog}
& \emph{Task \& domain} & Sound matching $[x]\to[\bfP]$. The synthesizer parameters $\bfP$ were estimated to match the reference (target) audio $x$. \\
\cmidrule{2-3}
& \emph{Processors} & Oscillators, envelope generators, and filters that allow parameter modulation as an optional input. \\
\cmidrule{2-3}
& \emph{Graph} & Any pre-defined directed acyclic graph (DAG). For example, a subtractive synthesizer that comprises $2$ oscillators, $1$ amplitude envelope, and $1$ lowpass filter were used in the experiments. \\
\cmidrule{2-3}
& \emph{Method} & Trained a single neural backbone for the reference encoding, followed by multiple prediction heads for the parameters. Optimized with a parameter loss and spectral loss, where the latter is calculated with every intermediate output. \\
\arrayrulecolor{black}
\midrule
\arrayrulecolor{lightgray}

\cite{caspe2022ddx7}
& \emph{Task \& domain} 
& Sound matching $[x]\to[\bfP]$. A frequency-modulation (FM) synthesizer matches recordings of monophonic instruments (violin, flute, and trumpet). 
Estimates parameters of an operator graph that is empirically searched \& selected. \\
\cmidrule{2-3}
& \emph{Processors} 
& Differentiable sinusoidal oscillators, each used as a carrier or modulator, pre-defined frequencies. 
An additional FIR reverb is added to the FM graph output for post-processing. \\
\cmidrule{2-3}
& \emph{Graph} 
& DAGs with at most $6$ operators. Different graphs for different target instruments.\\
\cmidrule{2-3}
& \emph{Method} 
& Trained a convolutional neural network that estimates envelopes from the target loudness and pitch. \\
\arrayrulecolor{black}
\midrule
\arrayrulecolor{lightgray}

\cite{ye2023fm}
& \emph{Task \& domain} 
& Sound matching $[x]\to [G, \bfP]$. Similar setup to the above \cite{caspe2022ddx7} plus additional estimation of the operator graph $G$. \\
\cmidrule{2-3}
& \emph{Processors} 
& Identical to \cite{caspe2022ddx7}, except for the frequency ratio that can be searched. \\
\cmidrule{2-3}
& \emph{Graph} 
& A subgraph of a supergraph, which resembles a multi-layer perceptron (modulator layers followed by a carrier layer). \\
\cmidrule{2-3}
& \emph{Method} 
& Trained a parameter estimator for the supergraph and found the appropriate subgraph $G$ with an evolutionary search. \\
\arrayrulecolor{black}
\midrule
\arrayrulecolor{lightgray}

\cite{Mitcheltree_2021}
& \emph{Task \& domain} 
& Reverse engineering $[s, y]\to [G, \bfP]$ of an audio effect chain from a subtractive synthesizer (commercial plugin). \\
\cmidrule{2-3}
& \emph{Processors} 
& $5$ audio effects: compressor, distortion, equalizer, phaser, and reverb. Non-differentiable implementations. \\
\cmidrule{2-3}
& \emph{Graph} 
& Chain of audio effects generated with no duplicate types (therefore $32$ possible combinations) and random order. \\ 
\cmidrule{2-3}
& \emph{Method} 
& Trained a next effect predictor and parameter estimator in a supervised (teacher-forcing) manner. \\
\arrayrulecolor{black}
\midrule
\arrayrulecolor{lightgray}

\cite{guo2023automatic}
& \emph{Task \& domain} 
& Blind estimation $[y]\to [G]$ and reverse engineering $[s,y]\to [G]$ of guitar effect chains.  \\
\cmidrule{2-3}
& \emph{Processors} 
& $13$ guitar effects, including non-linear processors, modulation effects, ambiance effects, and equalizer filters. \\
\cmidrule{2-3}
& \emph{Graph} 
& A chain of guitar effects. Maximum $5$ processors and a total of $221$ possible combinations. \\
\cmidrule{2-3}
& \emph{Method} 
& Trained a convolutional neural network with synthetic data to predict the correct combination. \\
\arrayrulecolor{black}
\midrule
\arrayrulecolor{lightgray}

\cite{steinmetz2020diffmixconsole} 
& \emph{Task \& domain} 
& Automatic mixing $[\bfS]\to[\bfP]$. Estimated parameters of fixed processing chains from source tracks $(K\leq 16)$. \\
\cmidrule{2-3}
& \emph{Processors} 
& $7$ differentiable processors, where $4$ (gain, polarity, fader, and panning) were implemented exactly. 
A combined effect of the remaining $3$ (equalizer, compressor, and reverb) was approximated with a single pre-trained neural network. \\
\cmidrule{2-3}
& \emph{Graph} 
& Tree structure: applied a fixed chain of the $7$ processors for each track, and then summed the chain outputs together. \\
\cmidrule{2-3}
& \emph{Method} 
& Trained a parameter estimator (convolutional neural network) with a spectrogram loss end-to-end. \\
\arrayrulecolor{black}
\midrule
\arrayrulecolor{lightgray}

\cite{ramirez2021differentiable} 
& \emph{Task \& domain} 
& Reverse engineering of music mastering $[s,y]\to[\bfP]$ \\
\cmidrule{2-3}
& \emph{Processors} 
& A multi-band compressor, graphic equalizer, and limiter. Gradient approximated with a finite difference method. \\
\cmidrule{2-3}
& \emph{Graph} 
& A serial chain of the processors. \\
\cmidrule{2-3}
& \emph{Method} 
& Optimized parameters with gradient descent. \\
\arrayrulecolor{black}
\midrule
\arrayrulecolor{lightgray}

\cite{lee2023blind}
& \emph{Task \& domain} 
& Blind estimation $[y]\to[G,\bfP]$ and reverse engineering $[\bfS, y]\to[G, \bfP]$.
Estimates both graph and its parameters for singing voice effect $(K=1)$ or drum mixing $(K\leq 6)$. \\
\cmidrule{2-3}
& \emph{Processors} 
& A total of $33$ processors, including linear filters, nonlinear filters, and control signal generators. 
Some processors are multiple-input multiple-output (MIMO), e.g., allowing auxiliary modulations. Non-differentiable implementations. \\ 
\cmidrule{2-3}
& \emph{Graph} 
& Complex DAG; splits (e.g., multi-band processing) and merges (e.g., sum and modulation). $30$ processors max. \\
\cmidrule{2-3}
& \emph{Method} 
& Trained a convolutional neural network-based reference encoder and a transformer variant for graph decoding and parameter estimation. 
Both were jointly trained via direct supervision of synthetic graphs (e.g., parameter loss). \\
\arrayrulecolor{black}
\midrule
\arrayrulecolor{lightgray}

\cite{colonel2023music} 
& \emph{Task \& domain} 
& Reverse engineering $[\bfS, y]\to [\bfP]$ of music mixing. Estimated parameters of a fixed chain for each track. \\
\cmidrule{2-3}
& \emph{Processors} 
& $6$ differentiable processors: gain, equalizer, compressor, distortion, panning, and reverb. \\
\cmidrule{2-3}
& \emph{Graph} & A chain of $5$ processors (all above types except the reverb) for each dry track (any other DAG can also be used). The reverb is used for the mixed sum. \\
\cmidrule{2-3}
& \emph{Method} & Parameters were optimized with spectrogram loss end-to-end via gradient descent. \\
\arrayrulecolor{black}
\midrule
\arrayrulecolor{lightgray}

Ours
& \emph{Task \& domain} 
& Reverse engineering $[\bfS, y]\to [G, \bfP]$ of music mixing. Estimated a chain of processors and their parameters for each track and submix where $K\leq 130$. \\
\cmidrule{2-3}
& \emph{Processors} & $7$ differentiable processors: gain/panning, stereo imager, equalizer, reverb, compressor, noisegate, and delay. \\
\cmidrule{2-3}
& \emph{Graph} & A tree of processing chains with a subgrouping structure (any other DAG can also be used). Processors can be omitted but should follow the fixed order. \\
\cmidrule{2-3}
& \emph{Method} & Joint estimation of the soft masks (dry/wet weights) and processor parameters.
Optimized with the spectrogram loss (and additional regularizations) end-to-end via gradient descent.
Accompanied by hard pruning stages. 
\\
\arrayrulecolor{black}

\bottomrule
\end{tabular}
\label{table:related_works}
\end{center}
\end{table*}

\clearpage
\newpage

\begin{table*}[t]
\setlength\tabcolsep{2.8pt}
\renewcommand{\arraystretch}{.8}
\begin{center}
\caption{
\it
Per-dataset results of the mixing consoles with different processor type configurations.
}
\begin{tabular}{lc|cccccccccccc}
\toprule
& & \multicolumn{4}{c}{\texttt{MedleyDB}} & \multicolumn{4}{c}{\texttt{MixingSecrets}} & \multicolumn{4}{c}{\texttt{Internal}}\\
\cmidrule(r){3-6}
\cmidrule(r){7-10}
\cmidrule{11-14}
 & 
 & $L_\mathrm{a}$ & $L_\mathrm{lr}$ & $L_\mathrm{m}$ & $L_\mathrm{s}$ 
 & $L_\mathrm{a}$ & $L_\mathrm{lr}$ & $L_\mathrm{m}$ & $L_\mathrm{s}$ 
 & $L_\mathrm{a}$ & $L_\mathrm{lr}$ & $L_\mathrm{m}$ & $L_\mathrm{s}$ 

\\ \midrule
Base graph 
& \texttt{\textcolor{white}{ecnigdr}}
& $50.7$ & $1.45$ & $1.42$ & $198$
& $7.30$ & $2.16$ & $2.02$ & $22.9$
& $1.12$ & $.951$ & $.940$ & $1.63$
\\ 
\midrule
+ Gain/panning 
& \texttt{\textcolor{white}{ecns\textcolor{black}{g}dr}}
& $.550$ & $.583$ & $.485$ & $.550$ 
& $.876$ & $.856$ & $.819$ & $.973$
& $.642$ & $.619$ & $.597$ & $.734$
\\ 
+ Stereo imager 
& \texttt{\textcolor{white}{ecn\textcolor{black}{sg}dr}} 
& $.541$ & $.564$ & $.483$ & $.553$
& $.847$ & $.834$ & $.791$ & $.928$
& $.538$ & $.616$ & $.595$ & $.727$
\\ 
+ Equalizer 
& \texttt{\textcolor{white}{\textcolor{black}{e}cn\textcolor{black}{sg}dr}} 
& $.450$ & $.453$ & $.390$ & $.504$
& $.700$ & $.698$ & $.622$ & $.780$
& $.522$ & $.497$ & $.467$ & $.626$
\\ 
+ Reverb 
& \texttt{\textcolor{white}{\textcolor{black}{e}cn\textcolor{black}{sg}d\textcolor{black}{r}}}
& $.368$ & $.361$ & $.360$ & $.390$ 
& $.614$ & $.601$ & $.579$ & $.674$
& $.463$ & $.451$ & $.432$ & $.517$
\\ 
+ Compressor 
& \texttt{\textcolor{white}{\textcolor{black}{ec}n\textcolor{black}{sg}d\textcolor{black}{r}}}
& $.315$ & $.304$ & $.297$ & $.356$
& $.558$ & $.542$ & $.512$ & $.637$
& $.396$ & $.377$ & $.347$ & $.482$
\\ 
+ Noisegate 
& \texttt{\textcolor{white}{\textcolor{black}{ecnsg}d\textcolor{black}{r}}}
& $.302$ & $.288$ & $.281$ & $.353$
& $.548$ & $.532$ & $.502$ & $.625$ 
& $.393$ & $.374$ & $.343$ & $.480$
\\ 
\midrule
+ Multitap delay (full) 
& \texttt{\textcolor{white}{\textcolor{black}{ecnsgdr}}}
& $.296$ & $.288$ & $.284$ & $.324$
& $.545$ & $.529$ & $.502$ & $.618$
& $.385$ & $.369$ & $.338$ & $.465$
\\ 
\bottomrule
\end{tabular}

\label{table:mixing-console-per-dataset}
\end{center}
\vspace{-3mm}
\end{table*}

\begin{table*}
\setlength\tabcolsep{2.3pt}
\renewcommand{\arraystretch}{.8}
\begin{center}
\caption{
\it
	Per-dataset results of the pruning. 
	\texttt{MC}: mixing console. \texttt{BF}: brute-force, \texttt{DW}: dry/wet, and \texttt{H}: hybrid.
}
\begin{tabular}{ll|ccccccccc ccccccccc ccccccccc}
\toprule
& & \multicolumn{9}{c}{\texttt{MedleyDB}} & \multicolumn{9}{c}{\texttt{MixingSecrets}} & \multicolumn{9}{c}{\texttt{Internal}} 
\\
\cmidrule(r){3-11}
\cmidrule(r){12-20}
\cmidrule{21-29}
 & \multicolumn{1}{c|}{$\tau$} & $L_\mathrm{a}$
 & $\rho$ 
 & $\rho_\textttt{g}$ 
 & $\rho_\textttt{s}$ 
 & $\rho_\textttt{e}$ 
 & $\rho_\textttt{r}$ 
 & $\rho_\textttt{c}$ 
 & $\rho_\textttt{n}$ 
 & $\rho_\textttt{d}$ 
 & $L_\mathrm{a}$
 & $\rho$ 
 & $\rho_\textttt{g}$ 
 & $\rho_\textttt{s}$ 
 & $\rho_\textttt{e}$ 
 & $\rho_\textttt{r}$ 
 & $\rho_\textttt{c}$ 
 & $\rho_\textttt{n}$ 
 & $\rho_\textttt{d}$
 & $L_\mathrm{a}$
 & $\rho$ 
 & $\rho_\textttt{g}$ 
 & $\rho_\textttt{s}$ 
 & $\rho_\textttt{e}$ 
 & $\rho_\textttt{r}$ 
 & $\rho_\textttt{c}$ 
 & $\rho_\textttt{n}$ 
 & $\rho_\textttt{d}$
 \\
\midrule

\texttt{MC}
& \multicolumn{1}{c|}{$-$}
& $.296$ & $-$ & $-$ & $-$ & $-$ & $-$ & $-$ & $-$ & $-$
& $.545$ & $-$ & $-$ & $-$ & $-$ & $-$ & $-$ & $-$ & $-$
& $.385$ & $-$ & $-$ & $-$ & $-$ & $-$ & $-$ & $-$ & $-$
\\

\midrule
\texttt{BF} & $.01$
& $.305$ & $.63$ & $.35$ & $.81$ & $.54$ & $.77$ & $.67$ & $.71$ & $.53$
& $.566$ & $.65$ & $.50$ & $.82$ & $.43$ & $.71$ & $.61$ & $.77$ & $.75$ 
& $.402$ & $.80$ & $.76$ & $.92$ & $.61$ & $.81$ & $.85$ & $.87$ & $.80$
\\

\midrule
\texttt{DW} & $.01$ 
& $.302$ & $.56$ & $.32$ & $.79$ & $.42$ & $.74$ & $.57$ & $.56$ & $.43$ 
& $.561$ & $.57$ & $.47$ & $.82$ & $.22$ & $.62$ & $.54$ & $.74$ & $.55$ 
& $.397$ & $.74$ & $.74$ & $.82$ & $.49$ & $.70$ & $.87$ & $.86$ & $.61$ 
\\
\midrule
\multirow{3}{*}{\texttt{H}} & $.001$ 
& $.295$ & $.44$ & $.22$ & $.73$ & $.26$ & $.61$ & $.50$ & $.54$ & $.24$
& $.550$ & $.43$ & $.35$ & $.76$ & $.13$ & $.42$ & $.43$ & $.55$ & $.38$ 
& $.388$ & $.59$ & $.48$ & $.77$ & $.40$ & $.57$ & $.78$ & $.77$ & $.39$ 
\\

& $.01$  
& $.302$ & $.61$ & $.32$ & $.81$ & $.48$ & $.74$ & $.69$ & $.71$ & $.52$ 
& $.563$ & $.62$ & $.49$ & $.85$ & $.34$ & $.64$ & $.60$ & $.79$ & $.65$
& $.400$ & $.77$ & $.73$ & $.93$ & $.56$ & $.75$ & $.85$ & $.86$ & $.74$
\\

& $.1$
& $.375$ & $.83$ & $.59$ & $.93$ & $.83$ & $.92$ & $.80$ & $.84$ & $.90$
& $.648$ & $.84$ & $.70$ & $.91$ & $.75$ & $.86$ & $.83$ & $.93$ & $.91$ 
& $.474$ & $.93$ & $.90$ & $.99$ & $.84$ & $.92$ & $.93$ & $.95$ & $.96$ 
\\

\bottomrule
\end{tabular}
\label{table:pruning-per-dataset}
\end{center}
\vspace{-4mm}
\end{table*}

\begin{algorithm}[H]
\caption{
\it
Music mixing graph search (dry/wet method).
}
\begin{algorithmic}[1]
\Require A mixing console $\Gc$, dry tracks $\bfS$, and mixture $y$
\Ensure Pruned graph $\Gp$, parameters $\bfP$, and weights $\bfw$
\vspace{.7mm}

\State $\bfP, \bfw \gets \fn{Initialize}(\Gc)$ 
\State $\bfP, \bfw \gets \fn{Train}(\Gc, \bfP, \bfw, \bfS, y)$
\State $\La^\mathrm{min} \gets \fn{Evaluate}(\Gc, \bfP, \bfw, \bfS, y)$ 

\State $\Gp \gets \Gc$

\For{$n$ $\leftarrow$ $1$ \textbf{to} $N_\mathrm{iter}$} 
	\color{blue}
	\State $\Tc \gets \fn{GetProcessorTypeSet}(V)$ \label{algo:prune-weight:init-start}
	\For{$t$ \textbf{in} $\Tc$}
		\State $V_t, \bfw_t \gets \fn{Filter}(V, t), \fn{Filter}(\bfw, t)$
		\State $N_t, r_t, \bfm_t \gets |V_t|, 0.1, \mathbf{1}$
	\EndFor \label{algo:prune-weight:init-end}
	\color{black}
	\color{purple}
	\While {$\Tc\neq\emptyset$} \label{algo:prune-weight:while}
	\color{red}
		\State $t \gets \fn{SampleType}(\Tc)$ \label{algo:prune-weight:sample-start}
		\State $\bar{V}_t, \bar{\bfm} \gets 
		\fn{GetLeastWeightNodes}(V_t, \bfw_t, \lfloor N_t r_t \rceil )$ \label{algo:prune-weight:sample-end}
		\color{black}
		\State $\La \gets \fn{Evaluate}(\Gp, \bfP, \bfw\odot \bfm \odot \bar{\bfm}, \bfS, y)$ 
		\If {$\La < \La^\mathrm{min} + \tau$} 
			\State $\La^\mathrm{min} \gets \min (\La^\mathrm{min}, \La) $ 
            \State $\bfm \gets \bfm \odot \bar{\bfm}$
			\color{magenta}
			\State $V_t \gets V_t \setminus \bar{V}_t$ \label{algo:prune-weight:update-success}
			\color{black}
		\Else
			\color{magenta}
			\If {$\lfloor N_t r_t \rceil > 1$} \label{algo:prune-weight:update-fail-start}
				\State $r_t \gets r_t/2$
			\Else
				\State $\Tc \gets \Tc \setminus \{t\}$
			\EndIf
			\color{black}
		\EndIf \label{algo:prune-weight:update-fail-end}
    \EndWhile
	\color{black}
	\State $\Gp, \bfP, \bfw \gets \fn{Prune}(\Gp, \bfP, \bfw, \bfm)$ 
    \State $\bfP, \bfw \gets \fn{Train}(\Gp, \bfP, \bfw, \bfS, y)$
\EndFor 
\State \Return $\Gp, \bfP, \bfw$
\end{algorithmic}
\label{algo:pruning-weight} 
\end{algorithm}

\begin{figure}[!t]
  \vspace{-1.5mm}
  \centering
  \includegraphics[width=.99\columnwidth]{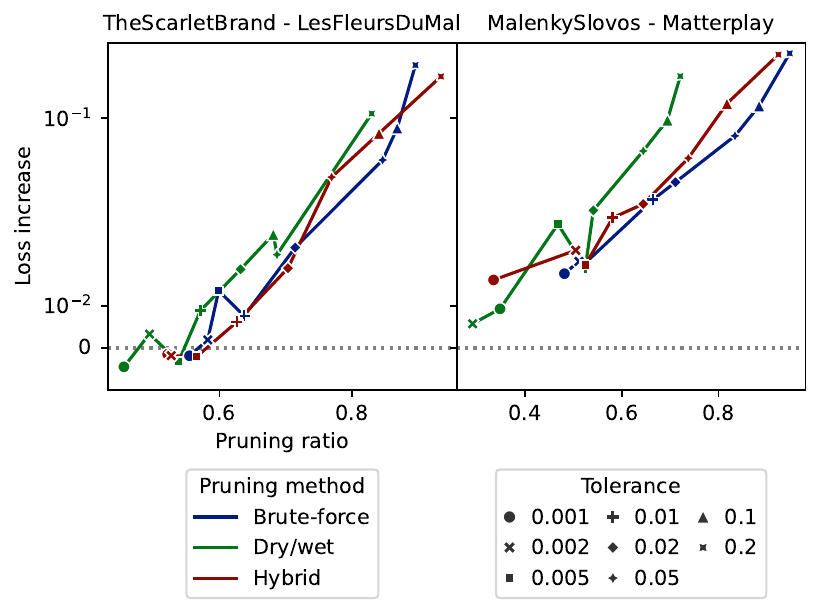}
  \vspace{-2mm}
  \caption{
  \it
	  	Loss increases from the mixing console and pruning ratios 
		for different pruning methods and tolerances.
  }
  \label{fig:loss-vs-ratio} 
  \vspace{-1mm}
\end{figure}

\section{Supplementary Results}\label{section:sup_results}
\begin{itemize}[leftmargin=4mm]
  \setlength\itemsep{0em}
    \item Table \ref{table:mixing-console-per-dataset} and \ref{table:pruning-per-dataset} report 
		the per-dataset results on the mixing consoles and graph pruning, respectively.
	\item Figure \ref{fig:loss-vs-ratio} compares the pruning methods on $2$ random-sampled songs
		using $7$ tolerance settings from $0.001$ to $0.2$.
	\item Figure \ref{fig:pruned-graphs-multiple} shows multiple graphs obtained by pruning the same console (song) repeatedly.
	\item Refer to Figure \ref{fig:pruned-graphs-1}-\ref{fig:pruned-graphs-3} for more pruned graphs 
		obtained with the default setting --- hybrid method and $\tau=0.01$.
	\item Figure \ref{fig:spec-medley}-\ref{fig:spec-internal} show more spectrogram plots.
 \end{itemize}

\begin{figure*}[th]
    \begin{center}
	\setlength\tabcolsep{2.9pt}
	\renewcommand{\arraystretch}{1}
	\begin{tabular}{lll}
    \subfloat{
        \includegraphics[height=.52\columnwidth]{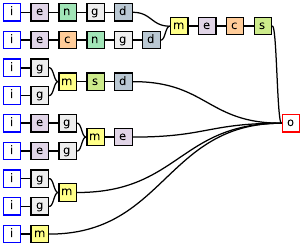}
		}
		&
    \subfloat{
        \includegraphics[height=.52\columnwidth]{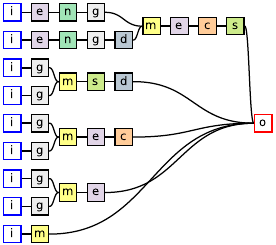}
		}
		&
    \subfloat{
        \includegraphics[height=.52\columnwidth]{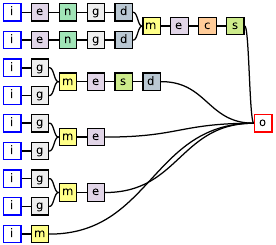}
		}
		\\
    \subfloat{
        \includegraphics[height=.52\columnwidth]{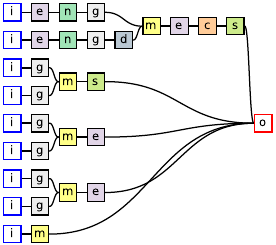}}
		&
    \subfloat{
        \includegraphics[height=.52\columnwidth]{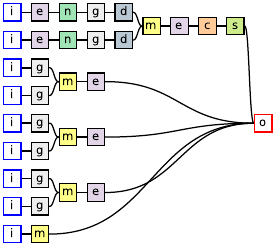}}
		&
    \subfloat{
        \includegraphics[height=.52\columnwidth]{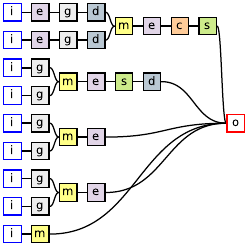}}
	\end{tabular}
    \captionsetup[subfloat]{captionskip=9pt}
    \vspace{-1mm}
	\caption{
 \it
	Each pruning run (default setting) yields a slightly different graph. 
	Song: \texttt{EthanHein\_GirlOnABridge}.
}
    \label{fig:pruned-graphs-multiple} 
    \end{center}
    \vspace{-2mm}
\end{figure*}

\begin{figure*}[th]
    \vspace{0mm}
    \begin{center}
    \captionsetup[subfloat]{captionskip=9pt}
    \subfloat[\it \texttt{Internal\_part1\_65536} \label{fig:pruned-a}]{
        \includegraphics[height=1.\columnwidth]{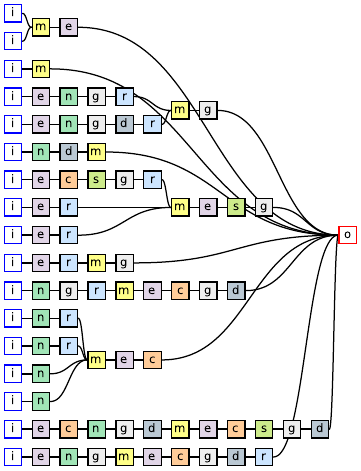}}
	\hspace{4mm}
    \subfloat[\it \texttt{DonCamilloChoir\_TrudeTheBumblebee} \label{fig:pruned-b}]{
        \includegraphics[height=1.180\columnwidth]{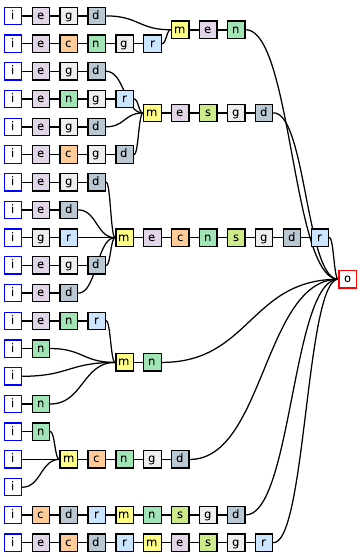}}
    \vspace{-1mm}
    \caption{\it Example pruned graphs (default setting). the number of tracks: $K \leq 20$.}
    \label{fig:pruned-graphs-1} 
    \end{center}
    \vspace{-3mm}
\end{figure*}

\begin{figure*}[th]
    \vspace{0mm}
    \begin{center}
    \captionsetup[subfloat]{captionskip=7pt}
    \subfloat[\it \texttt{LittleTybee\_TheAlchemist} \label{fig:pruned-c}]{
        \includegraphics[height=2.39\columnwidth]{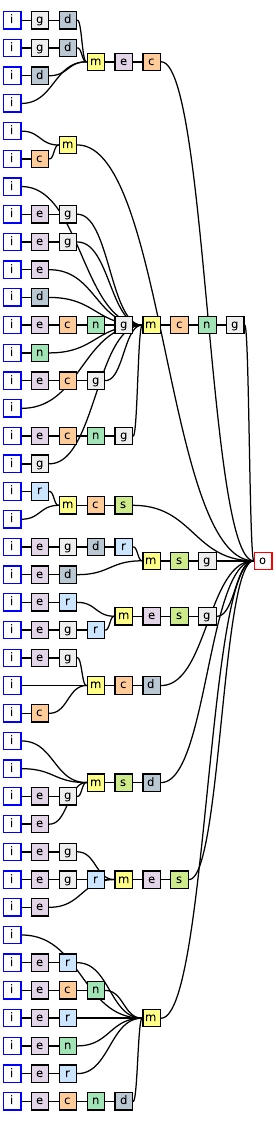}} 
	\hspace{4mm}
    \subfloat[\it \texttt{RaftMonk\_Tiring} \label{fig:pruned-d}]{
        \includegraphics[height=2.145\columnwidth]{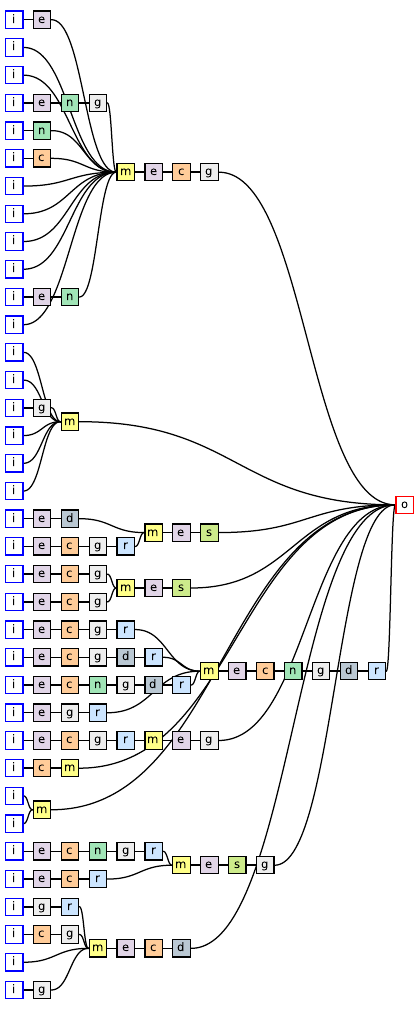}}
    \vspace{-1mm}
	\caption{\it Example pruned graphs (default setting). the number of tracks: $K > 20$.}
    \label{fig:pruned-graphs-2} 
    \end{center}
    \vspace{-3mm}
\end{figure*}

\begin{figure*}[th]
    \vspace{0mm}
    \begin{center}
    \captionsetup[subfloat]{captionskip=7pt}
    \subfloat[\it \texttt{Internal\_part2\_67692} \label{fig:pruned-e}]{
        \includegraphics[height=2.55\columnwidth]{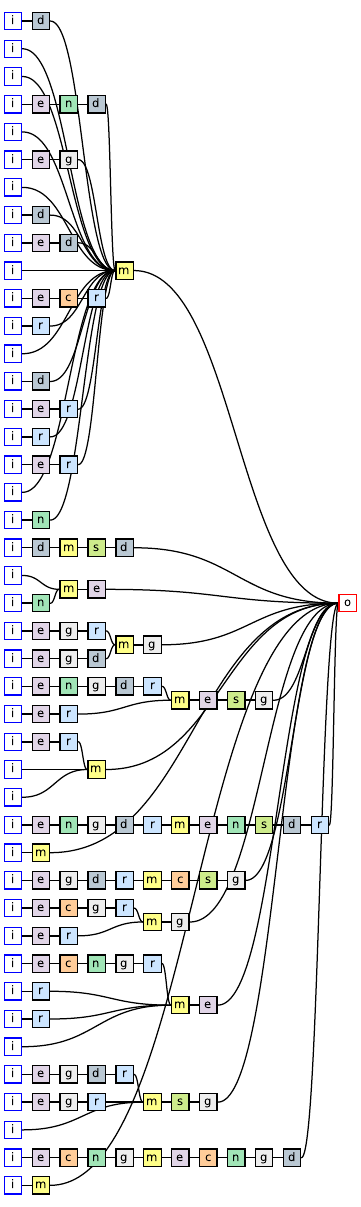}} 
	\hspace{4mm}
    \subfloat[\it \texttt{StevenClark\_Bounty} \label{fig:pruned-f}]{
        \includegraphics[height=2.436\columnwidth]{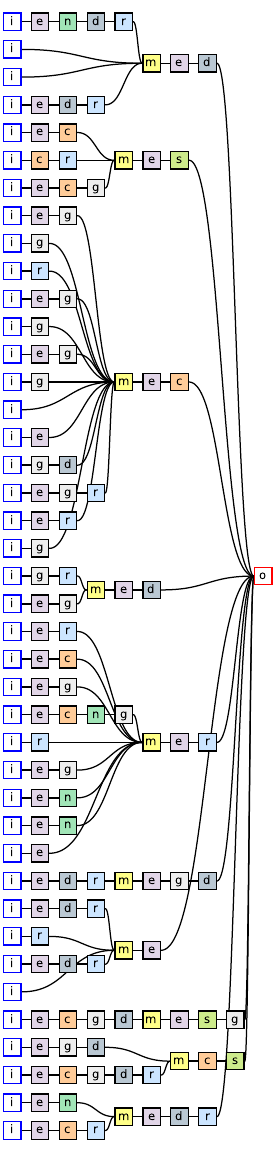}}
    \vspace{-1mm}
	\caption{\it Example pruned graphs (default setting). the number of tracks: $K > 20$ (continued).}
    \label{fig:pruned-graphs-3} 
    \end{center}
    \vspace{-3mm}
\end{figure*}

\begin{figure*}
  \centering
    \subfloat[\it \texttt{Torres\_NewSkin} \label{fig:spec-main-full-a}]{
        \includegraphics[width=.92\columnwidth]{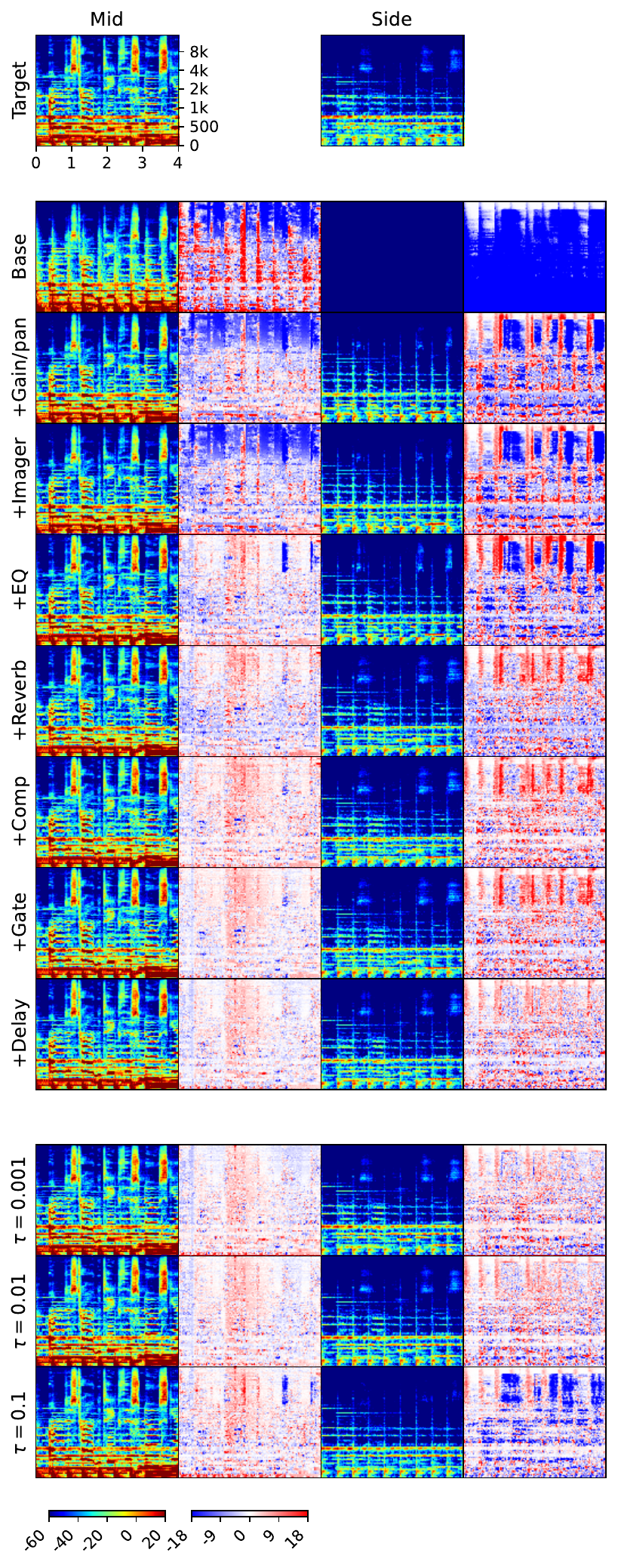}} \hspace{7mm}
    \subfloat[\it \texttt{TablaBreakbeatScience\_RockSteady} \label{fig:spec-main-full-b}]{
        \includegraphics[width=.92\columnwidth]{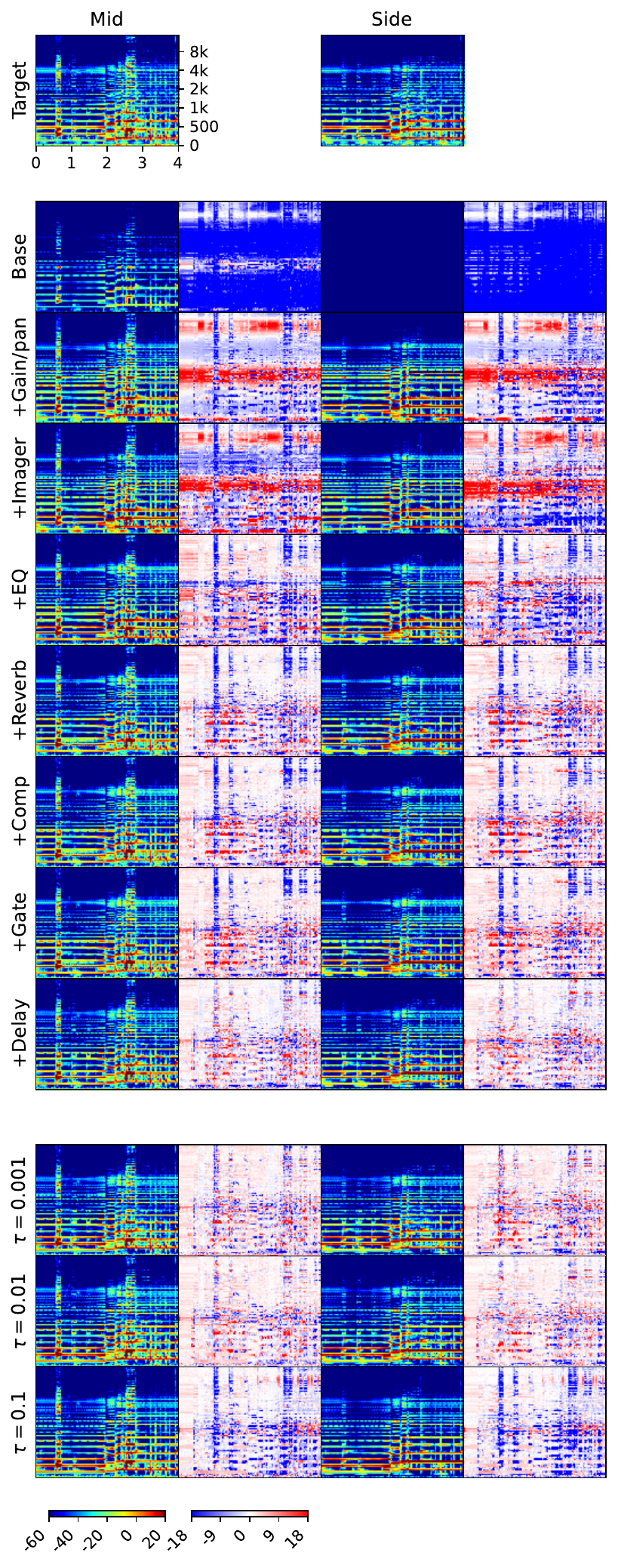}}\\
  \caption{\it Matching of target music mixes with mixing consoles and their pruned versions: \texttt{MedleyDB} dataset. 
  }
    \label{fig:spec-main-full} 
\end{figure*}

\begin{figure*}
  \centering
    \subfloat[\it \texttt{MusicDelta\_SwingJazz} \label{fig:spec-medley-a}]{
        \includegraphics[width=.92\columnwidth]{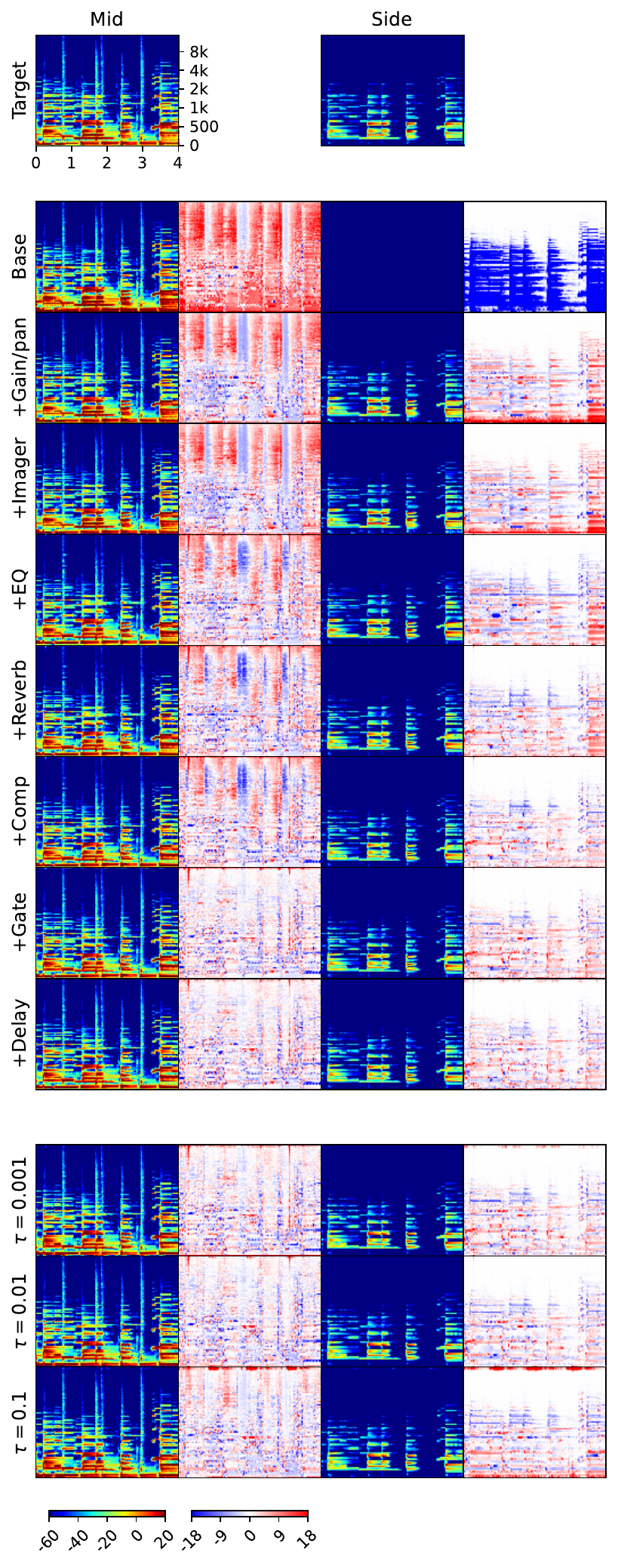}}
		\hspace{7mm}
    \subfloat[\it \texttt{ChrisJacoby\_BoothShotLincoln} \label{fig:spec-medley-b}]{
        \includegraphics[width=.92\columnwidth]{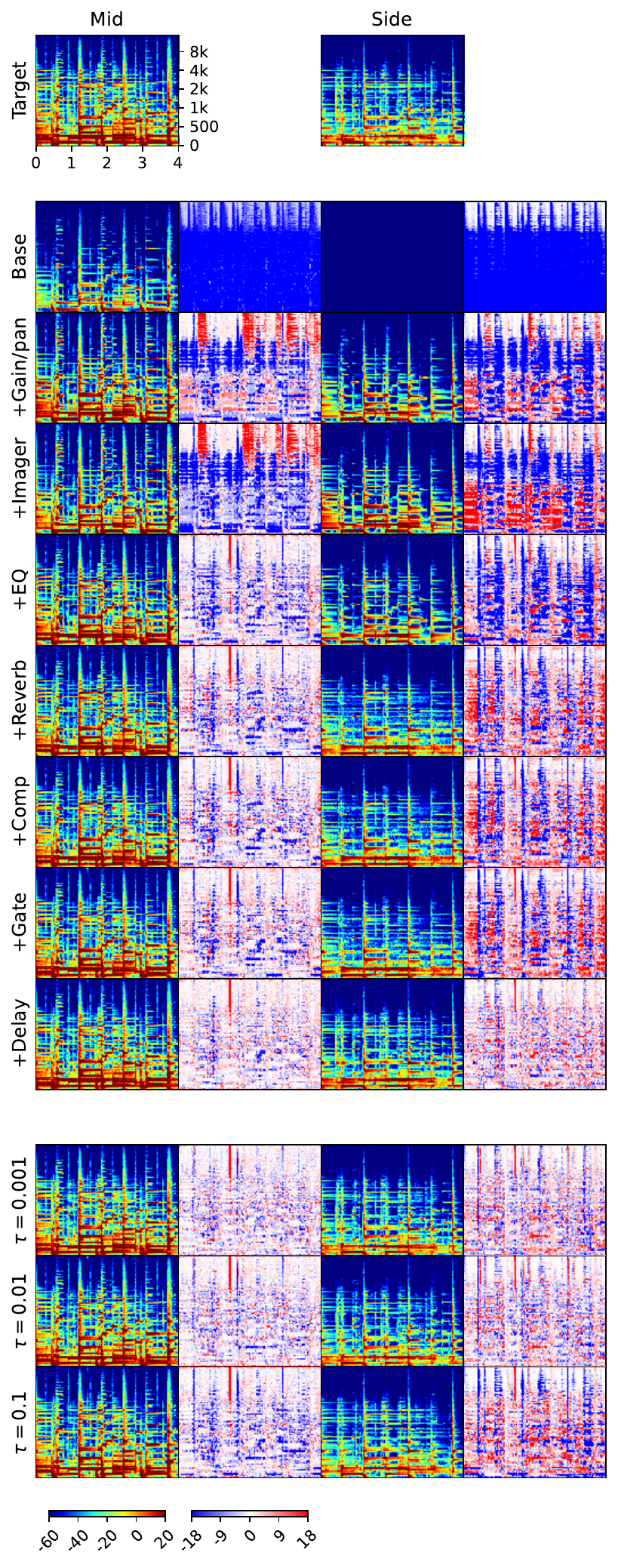}} 
		\\
	\caption{\it Matching of target music mixes with mixing consoles and their pruned versions: \texttt{MedleyDB} dataset. 
  }
    \label{fig:spec-medley} 
\end{figure*}

\begin{figure*}
  \centering
    \subfloat[\it \texttt{HowlProject\_IfIWereABell} \label{fig:spec-mixingsecrets-a}]{
        \includegraphics[width=.92\columnwidth]{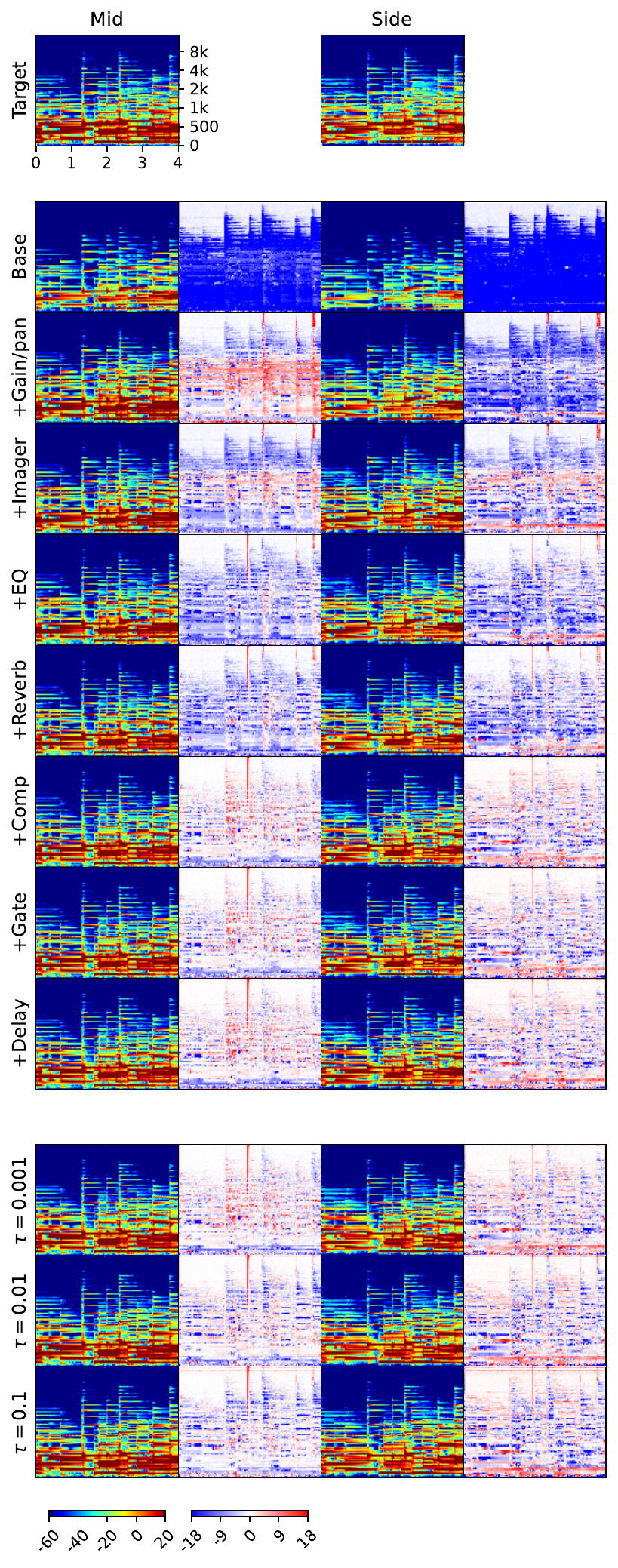}}
		\hspace{7mm}
    \subfloat[\it \texttt{IanDearden\_TeraniaCreekWalking} \label{fig:spec-mixingsecrets-b}]{
        \includegraphics[width=.92\columnwidth]{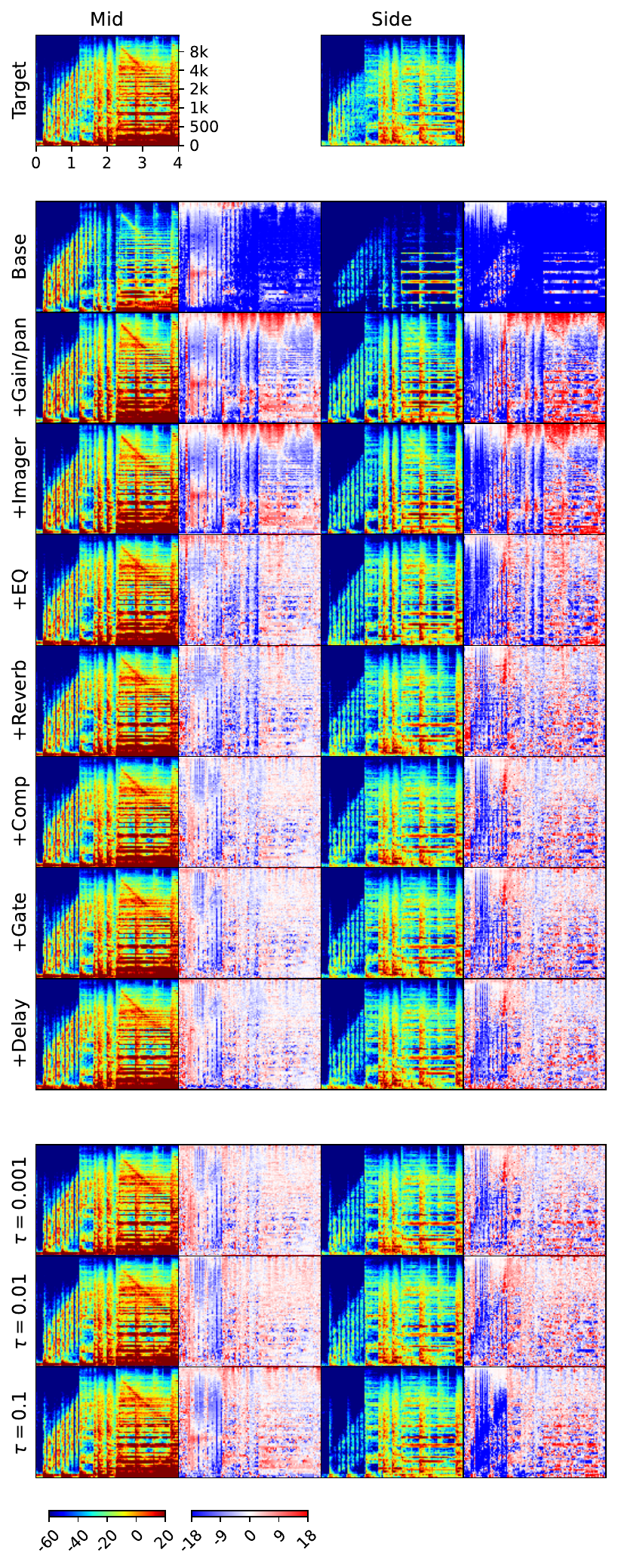}} 
		\\
	\caption{\it Matching of target music mixes with mixing consoles and their pruned versions:  \texttt{MixingSecrets} dataset. 
  }
    \label{fig:spec-mixingsecrets} 
\end{figure*}

\begin{figure*}
  \centering
    \subfloat[\it \texttt{Internal\_66680} \label{fig:spec-internal-a}]{
        \includegraphics[width=.92\columnwidth]{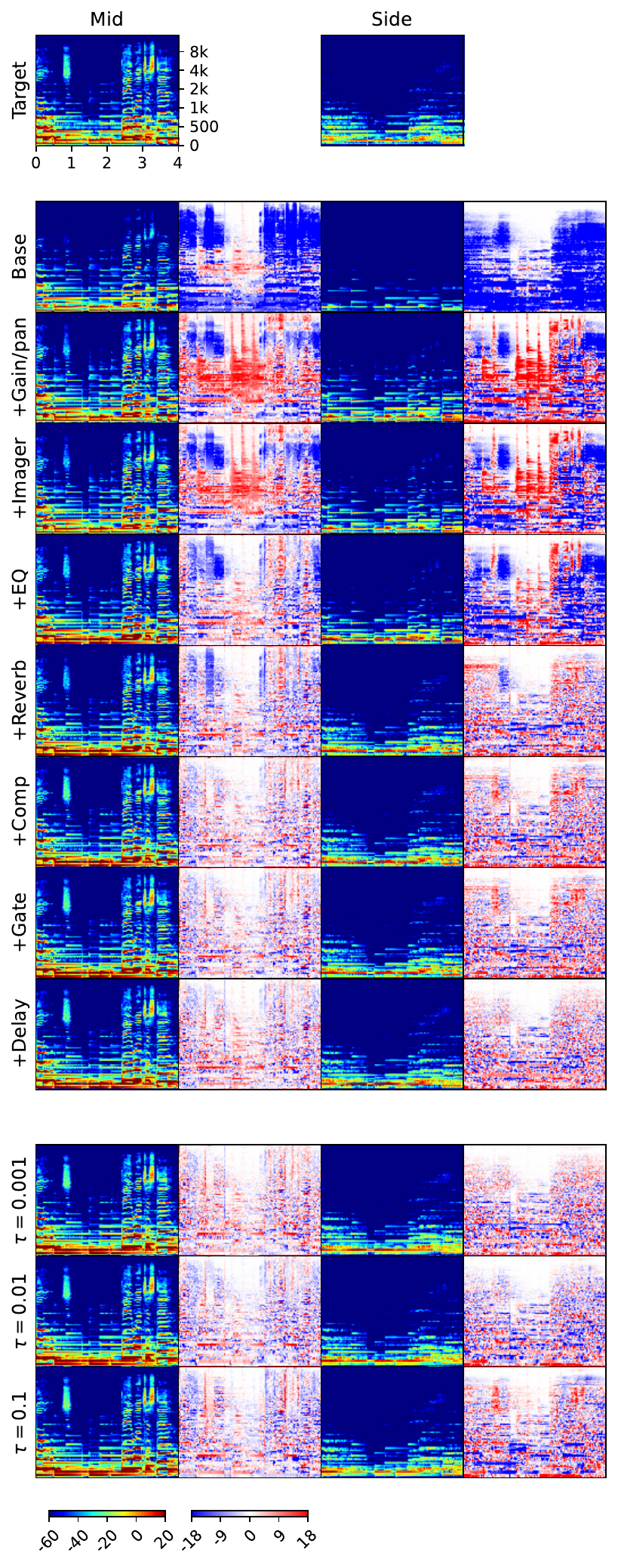}}
		\hspace{7mm}
    \subfloat[\it \texttt{Internal\_67954} \label{fig:spec-internal-b}]{
        \includegraphics[width=.92\columnwidth]{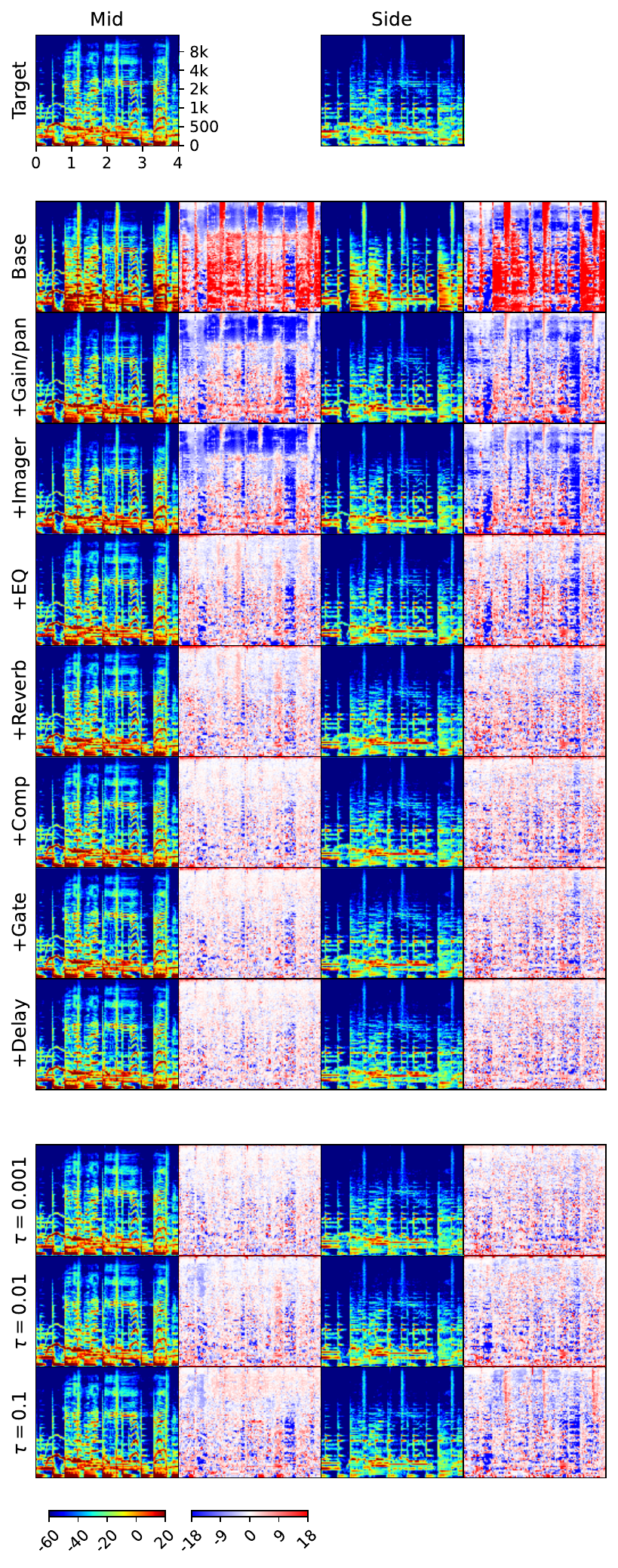}} 
		\\
	\caption{\it Matching of target music mixes with mixing consoles and their pruned versions: \texttt{Internal} dataset. 
  }
    \label{fig:spec-internal} 
\end{figure*}

\clearpage
\clearpage
\newpage
\bibliographystyle{IEEEbib}
\bibliography{refs} 
\fi

\iffull
\maketitle
\begin{abstract}
Music mixing is \emph{compositional} --- experts combine multiple audio processors to achieve a cohesive mix from dry source tracks. We propose a method to reverse engineer this process from the input and output audio. 
First, we create a mixing console that applies all available processors to every chain. Then, after the initial console parameter optimization, we alternate between removing redundant processors and fine-tuning. 
We achieve this through differentiable implementation of both processors and pruning.
Consequently, we find a sparse mixing graph that achieves nearly identical matching quality of the full mixing console. 
We apply this procedure to dry-mix pairs from various datasets and collect graphs that also can be used to train neural networks for music mixing applications.
\end{abstract}

\clearpage
\bibliographystyle{IEEEbib}
\bibliography{refs} 
\clearpage
\appendix

\fi

\end{document}